\documentclass[twocolumn,showpacs,aps,prd,nofootinbib]{revtex4}
\usepackage{graphicx}
\usepackage{dcolumn}
\usepackage{amsmath}
\usepackage{epsfig}
\usepackage{colordvi}
\usepackage{color}
\usepackage{hhline}

\RequirePackage{xspace}

\newcommand{\BABARPubYear}    {15}
\newcommand{\BABARPubNumber}  {004}
\newcommand{\SLACPubNumber} {16333}

\usepackage{relsize}
\def\babar{\mbox{\slshape B\kern-0.1em{\smaller A}\kern-0.1em
    B\kern-0.1em{\smaller A\kern-0.2em R}}}
    
\mathchardef\Upsilon="7107
\def\Y#1S{\ensuremath{\Upsilon{(#1S)}}\xspace}

\def\pep2{PEP-II}

\long\def\inst#1{\par\nobreak\kern 4pt\nobreak
  {\it #1}\par\vskip 10pt plus 3pt minus 3pt}
  
\begin{document}

\begin{flushleft}
\babar-PUB-\BABARPubYear/\BABARPubNumber \\
SLAC-PUB-\SLACPubNumber 
\end{flushleft}

\title{\large \bf
\boldmath
Study of the $e^+e^-\to K^+K^-$ reaction
in the energy range from 2.6 to 8.0 GeV}

%
\author{J.~P.~Lees}
\author{V.~Poireau}
\author{V.~Tisserand}
\affiliation{Laboratoire d'Annecy-le-Vieux de Physique des Particules (LAPP), Universit\'e de Savoie, CNRS/IN2P3,  F-74941 Annecy-Le-Vieux, France}
\author{E.~Grauges}
\affiliation{Universitat de Barcelona, Facultat de Fisica, Departament ECM, E-08028 Barcelona, Spain }
\author{A.~Palano$^{ab}$ }
\affiliation{INFN Sezione di Bari$^{a}$; Dipartimento di Fisica, Universit\`a di Bari$^{b}$, I-70126 Bari, Italy }
\author{G.~Eigen}
\author{B.~Stugu}
\affiliation{University of Bergen, Institute of Physics, N-5007 Bergen, Norway }
\author{D.~N.~Brown}
\author{L.~T.~Kerth}
\author{Yu.~G.~Kolomensky}
\author{M.~J.~Lee}
\author{G.~Lynch}
\affiliation{Lawrence Berkeley National Laboratory and University of California, Berkeley, California 94720, USA }
% ...
\collaboration{The \babar\ Collaboration}
\noaffiliation

\begin{abstract}
The $e^+e^-\to K^+K^-$ cross section and charged-kaon  
electromagnetic form factor are measured in the   
$e^+e^-$ center-of-mass energy range ($E$) from 2.6 to 8.0 GeV 
using the initial-state radiation technique with an undetected photon.
The study is performed using 469 fb$^{-1}$ of data collected
with the \babar\ detector at the \pep2\ $e^+e^-$ collider
at center-of-mass energies near 10.6 GeV.
The form factor is found to decrease with energy
faster than $1/E^2$, and  approaches the asymptotic QCD prediction.
Production of the $K^+K^-$ final state through the $J/\psi$
and $\psi(2S)$ intermediate states is observed.      
The results for the kaon form factor are used
together with data from other experiments to perform
a model-independent determination
of the relative phases between electromagnetic (single-photon)
and strong  
amplitudes in $J/\psi$ and $\psi(2S)\to K^+K^-$ decays.
The values of the branching fractions measured in the reaction 
$e^+e^- \to K^+K^-$ are shifted relative to their true values
due to interference between resonant and nonresonant
amplitudes. The values of these shifts are determined to be about 
$\pm5\%$ for the $J/\psi$ meson and $\pm15\%$ for the $\psi(2S)$ meson.
\end{abstract}


\maketitle

\setcounter{footnote}{0}

\section{\boldmath Introduction\label{intro}}
The timelike charged-kaon form factor $F_K$ has been measured precisely in 
the threshold/$\phi$-meson region~\cite{sndphi,cmdphi,KKbabar} and by several 
experiments~\cite{olya,dm1,dm2,sndhigh,KKbabar} in the center-of-mass 
(c.m.) energy range 
1.1--2.4 GeV, where substantial structure is evident.  At higher energies, 
there are precise measurements at 3.671 GeV~\cite{CLEO}, 3.772, and 
4.170 GeV~\cite{NU}, and there is a scan that extends to 5 GeV~\cite{KKbabar}. 
The energy dependence of these higher-energy data is consistent with the 
asymptotic form predicted by perturbative quantum chromodynamics (pQCD), 
but their magnitude is about a factor of four   
higher than the predicted asymptotic value~\cite{QCDpion}
\begin{equation}
{M_{K^+K^-}^2}|F_K(M_{K^+K^-})|=8\pi\alpha_s f_K^2,
\label{eqqcd}
\end{equation}
where $M_{K^+K^-}$ is the $K^+K^-$ invariant mass, 
$\alpha_s$ is the strong coupling constant, and $f_K=156.2\pm 0.7$ 
MeV~\cite[p.~1027]{pdg} is the charged-kaon decay constant\footnote{
We note that this value is larger by a factor of $\sqrt{2}$
than that used in Eq.~(22) of Ref.~\cite{KKbabar}.}.
It is expected that the difference between the data and the asymptotic
QCD prediction will decrease with increasing energy.
Precise measurements at higher energies are needed to
test this expectation.

In this paper we analyze the initial-state radiation (ISR) process 
$e^+e^-\to K^+K^-\gamma$. The $K^+K^-$ mass spectrum measured in this process
is related to the cross section of the nonradiative process $e^+e^-\to K^+K^-$.
Our previous measurement of $F_K$~\cite{KKbabar} used the ``large-angle'' 
(LA) ISR technique, in which the radiated photon is 
detected and the $e^+e^-\to K^+K^-\gamma$ event is fully reconstructed. 
This gives good precision near threshold, but the cross section decreases 
rapidly
with increasing energy, limiting that measurement to energies below 5 GeV. 
In this paper we utilize small-angle (SA) ISR events, in which the ISR 
photon is emitted close to the $e^+e^-$ collision axis, and so is undetected. 
This allows us to perform an independent and complementary measurement of the
charged-kaon form factor, which has better precision in the range 2.6--5 GeV,
and extends the measurements up to 8~GeV.

The Born cross section for the ISR process integrated over the kaon momenta
and the photon polar angle is 
\begin{multline}
\frac{{\rm d}\sigma_{K^+K^-\gamma}(M_{K^+K^-})}
{{\rm d}M_{K^+K^-}} =\\ 
\frac{2M_{K^+K^-}}{s}\, W(s,x)\,\sigma_{K^+K^-}(M_{K^+K^-}),
\label{eq1}
\end{multline}
where
$s$ is the $e^+e^-$ c.m. energy squared,
$x\equiv{2E_{\gamma}^\ast}/\sqrt{s}=1-M_{K^+K^-}^2/{s}$, and $E_{\gamma}^\ast$ 
is the ISR photon energy in the $e^+e^-$ c.m. frame\footnote{Throughout 
this paper, an asterisk denotes a quantity that is evaluated in 
the $e^+e^-$ c.m. frame, while quantities
without asterisks are evaluated in the laboratory frame.}.
The function $W(s,x)$, describing the probability for single ISR emission
at lowest-order in quantum electrodynamics, is known to
an accuracy better than 0.5\%~\cite{radf1,radf2,radf3}.
The $e^+e^-\to K^+K^-$ cross section is given in terms of the form factor by 
\begin{equation}
\sigma_{K^+K^-}(M_{K^+K^-}) = \frac{\pi\alpha^{2}\beta^3 C}{3M_{K^+K^-}^2}
|F_K(M_{K^+K^-})|^{2},
\label{eq4}
\end{equation}
where $\alpha$ is the fine-structure constant, 
$\beta =\sqrt{1-4m_K^2/M_{K^+K^-}^2}$, and $C$ is the final-state correction,
which, in particular, takes into account
extra photon radiation from the final state (see, e.g., Ref.~\cite{FSC}).
In the mass region under study the factor $C$ is close to unity,
and varies from 1.008 at 2.6 GeV/$c^2$ to 1.007 at 8~GeV/$c^2$.

In addition to the form factor, we measure the branching
fractions for the
decays  $J/\psi\to K^+K^-$ and $\psi(2S)\to K^+K^-$. 
For the latter we study the interference between 
the resonant and nonresonant $e^+e^-\to K^+ K^-$ amplitudes, and between
the single-photon and strong $\psi\to K\bar{K}$ amplitudes
(with $\psi=J/\psi$, $\psi(2S)$).
As a result, we extract the interference corrections to the
$J/\psi\to K^+K^-$ and $\psi(2S)\to K^+K^-$ branching fractions, which 
were not taken into account in previous measurements, and
determine the values of the phase difference between 
the single-photon and strong amplitudes in $J/\psi\to K\bar{K}$ and 
$\psi(2S)\to K\bar{K}$ decays. 
In contrast to previous determinations of this phase~\cite{ph1,ph2,NUjpsi},
we use a model-independent approach, calculating the single-photon decay 
amplitudes from our data on the charged-kaon form factor.

\section{ \boldmath The \babar\ detector, data, and simulated samples}
\label{detector}
We analyze a data sample corresponding to an integrated luminosity of
469~fb$^{-1}$~\cite{lum}
recorded with the  \babar\ detector at the SLAC \pep2\ 
asymmetric-energy (9-GeV $e^-$ and 3.1-GeV $e^+$) collider.
About 90\% of the data were collected at an $e^+e^-$ c.m.~energy of 10.58~GeV
(the $\Upsilon$(4S) mass), and 10\% at 10.54 GeV.

The \babar\ detector is described in detail elsewhere~\cite{ref:babar-nim}.
Charged-particle tracking is
provided by a five-layer silicon vertex tracker (SVT) and
a 40-layer drift chamber (DCH), operating in the 1.5 T magnetic field
of a superconducting solenoid.  The position and energy of a photon-produced
cluster are measured with a CsI(Tl) electromagnetic calorimeter (EMC).
Charged-particle identification (PID) is provided by specific ionization
(${\rm d}E/{\rm d}x$) measurements in the SVT and DCH, and by an internally reflecting
ring-imaging Cherenkov detector (DIRC). Muons are identified in
the solenoid's instrumented flux return (IFR).

Simulated samples of signal events, and background 
$e^+e^-\to \pi^+\pi^-\gamma$ and $\mu^+\mu^-\gamma$ events, are generated 
with the Phokhara~\cite{phokhara} Monte Carlo (MC) event generator, which 
takes into account next-to-leading order radiative 
corrections. To obtain realistic estimates for the pion and kaon cross 
sections,
the experimental values of the pion and kaon electromagnetic form 
factors measured in the CLEO experiment at $\sqrt{s}=3.67$ GeV~\cite{CLEO} 
are used in the event generator. The mass dependence of the form factors is 
assumed to be $1/m^2$, as predicted by asymptotic QCD ~\cite{QCDpion}. The 
process $e^+e^-\to e^+e^-\gamma$ is simulated with the BHWIDE
event generator~\cite{BHWIDE}. 

Two-photon background from the process $e^+e^-\to e^+e^-K^+K^-$ is 
simulated with the GamGam event generator~\cite{gamgam}.  
Background contributions from $e^+e^-\to q\bar{q}(\gamma_{\rm ISR})$, 
where $q$ represents a $u$, $d$, $s$ or $c$ quark, are simulated with the 
JETSET event generator~\cite{JETSET}. 

The detector response is simulated using the Geant4 package~\cite{GEANT4}.
The simulation takes into account the variations in the detector and
beam-background conditions over the running period of the experiment.

\section{Event selection\label{sel}}
We select events with two tracks of opposite charge originating from
the interaction region. The tracks must lie in the polar angle range 
$25.8^\circ <\theta < 137.5^\circ$ and be identified as kaons. The selected 
kaon candidates are fitted to a common vertex with a beam-spot constraint. The
$\chi^2$ probability for this fit is required to be greater than 0.1\%.
\begin{figure}
\includegraphics[width=.4\textwidth]{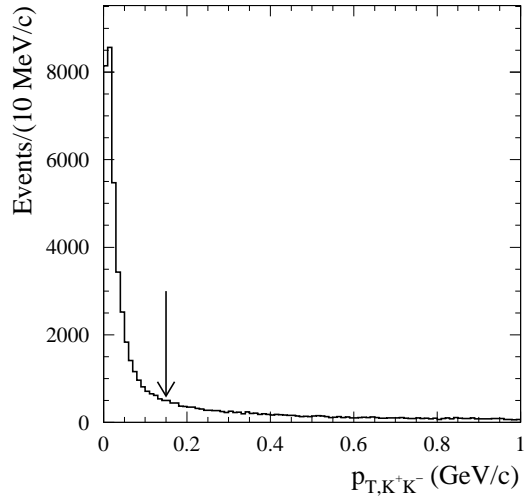}
\caption{The $p_{{\rm T},K^+K^-}$ distribution 
for simulated $e^+e^-\to K^+K^-\gamma$ events.
The arrow indicates $p_{{\rm T},K^+K^-}=0.15$ GeV/$c$.
\label{fig3}}
\end{figure}
\begin{figure}
\includegraphics[width=.4\textwidth]{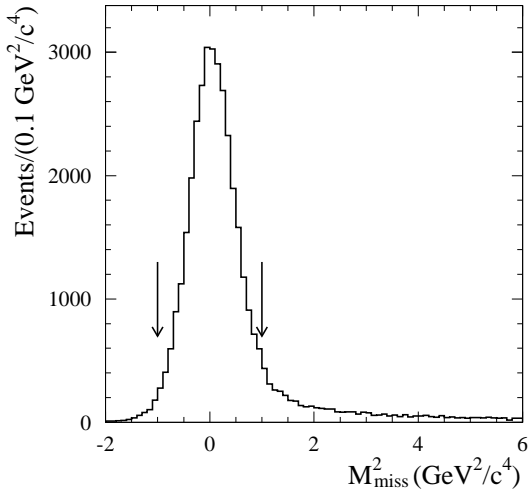}
\caption{The $M_{\rm{miss}}^2$ distribution
for simulated $e^+e^-\to K^+K^-\gamma$ events.
The arrows indicate $|M_{\rm{miss}}^2|=1$ GeV$^2/c^4$.\label{fig4}}
\end{figure}

Conditions on the $K^+K^-$ transverse
momentum ($p_{{\rm T},K^+K^-}$) and the missing-mass 
squared ($M_{\rm{miss}}^2$)
recoiling against the $K^+K^-$ system are used for further selection.
The $p_{{\rm T},K^+K^-}$ distribution
for simulated $e^+e^-\to K^+K^-\gamma$ events is shown in Fig.~\ref{fig3}.
The peak near zero corresponds to ISR photons emitted along
the collision axis, while the long tail is due to photons emitted at 
large angles. We apply the condition $p_{{\rm T},K^+K^-}<0.15$ GeV/$c$, which 
removes large-angle ISR and suppresses backgrounds from 
$e^+e^-\to K^+K^-\pi^0$ and ISR processes with extra $\pi^0$ mesons. 

The region of low $K^+K^-$ invariant mass  cannot 
be studied with small-angle ISR due to limited detector acceptance. 
A $K^+K^-$ pair with $p_{{\rm T},K^+K^-}<0.15$ GeV/$c$ is detected in \babar\  
when its invariant mass is larger than 2.5 (4.2) GeV/$c^2$ for 
an ISR photon emitted along the electron (positron) beam direction. 
The average values of the kaon momentum for the two photon directions 
are about 2.5 and 5 GeV/$c$, respectively. 
Since the probability for  particle misidentification increases strongly 
with increasing momentum, we reject events with an ISR photon 
along the positron direction.

The $M_{\rm{miss}}^2$ distribution for simulated signal events is shown in 
Fig.~\ref{fig4}. The signal distribution is peaked at zero, while
the background distributions are shifted to negative values
for $e^+e^-\to e^+e^-\gamma$ and $\mu^+\mu^-\gamma$ events and to positive 
values for $p\bar{p}\gamma$, two-photon and other ISR events.
The condition $|M_{\rm{miss}}^2|<1$ GeV$^2/c^4$ is 
applied to suppress background.
Sideband regions in $M_{\rm{miss}}^2$ and 
$p_{{\rm T},K^+K^-}$ are used to estimate the remaining 
background from these sources, as described in Sec.~\ref{background}.
\begin{figure}
\includegraphics[width=.4\textwidth]{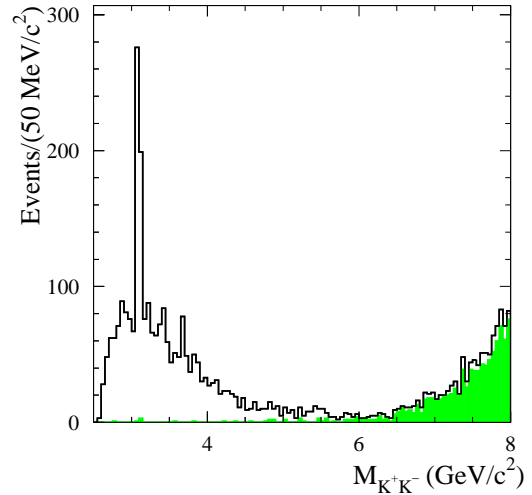}
\caption{The $K^+K^-$ mass spectrum for
selected $K^+K^-\gamma$ candidates. The peaks near 
3.1, 3.4, and 3.7 GeV$/c^2$ are from $J/\psi$, $\chi_{c0}$, and $\psi(2S)$
decays to $K^+K^-$, respectively. The shaded histogram shows
events with at least one identified muon candidate.
\label{fig5}}
\end{figure}

The $K^+K^-$ invariant-mass spectrum for events selected with the criteria
described above is shown in Fig.~\ref{fig5}. A clear $J/\psi$ signal
is seen in the spectrum, and there are also
indications of small $\psi(2S)$ and $\chi_{c0}$ peaks.
The $\chi_{c0}$ mesons are produced in the reaction 
$e^+e^-\to \psi(2S)\gamma \to \chi_{c0}\gamma\gamma$. The increase
in the number of events for $M_{K^+K^-}>6$ GeV/$c^2$ is
due to background from the $e^+e^-\to \mu^+\mu^-\gamma$ process.
To suppress the muon background we apply the additional condition 
that neither kaon candidate be identified as a muon.
Muon identification is based mainly on IFR information, and
does not make use of the DIRC and ${\rm d}E/{\rm d}x$ measurements used
for charged-kaon PID.
 For $\mu^+\mu^-\gamma$ background events,
the probability for at least one of the charged particles to be 
identified as a muon is about 88\% (see Subsection~\ref{bkg_mumu}).
The shaded histogram in Fig.~\ref{fig5} shows events with at least one 
identified muon candidate. The large muon background 
for larger values of $M_{K^+K^-}$
prevents us from providing results for $M_{K^+K^-}>8$ GeV/$c^2$.

The mass spectrum with finer binning in the region of
the charmonium resonances (3.0--4.5 GeV/$c^2$) is presented in the
Supplementary Material at [URL will be inserted by publisher], 
together with the mass resolution functions
obtained from MC simulation with $M_{K^+K^-}$ near the 
$J/\psi$ and $\psi(2S)$ masses.

\section{Background estimation and subtraction}\label{background}
Sources of background in the selected sample are:
other two-body ISR processes $e^+e^-\to e^+e^-\gamma$, $\mu^+\mu^-\gamma$,
$\pi^+\pi^-\gamma$, and $p\bar{p}\gamma$;
ISR processes containing additional neutral particles, e.g., 
$e^+e^-\to K^+K^- \pi^0\gamma$ and $e^+e^- \to \psi(2S) \gamma \to
\chi_{cJ} \gamma\gamma \to K^+K^-\gamma \gamma$;
the two-photon process $e^+e^-\to e^+e^-K^+K^-$; and
non-radiative $e^+e^-\to q\bar{q}$ events containing a $K^+K^-$ pair plus 
neutrals, e.g., $e^+e^-\to K^+K^- \pi^0$.
The background from the process $e^+e^-\to K^+K^- \pi^0$, which was 
dominant in our LA analysis~\cite{KKbabar}, is strongly 
suppressed by the requirement on $p_{{\rm T},K^+K^-}$,
and is found to be negligible in the SA analysis.
The cross section for 
$e^+e^-\to p\bar{p}\gamma$~\cite{ppbar1,ppbar2} is smaller than that for 
$e^+e^-\to K^+K^-\gamma$ in the mass region of interest, 
and this background is reduced to a negligible level 
by the requirement $|M_{\rm{miss}}^2|<1$ GeV$^2/c^4$.
The other categories of background are discussed in the 
following subsections. 
\subsection{ \boldmath Background from $e^+e^-\to e^+e^-\gamma$, 
$\mu^+\mu^-\gamma$, and $\pi^+\pi^-\gamma$\label{bkg_mumu}}
\begin{figure*}
\includegraphics[width=.32\textwidth]{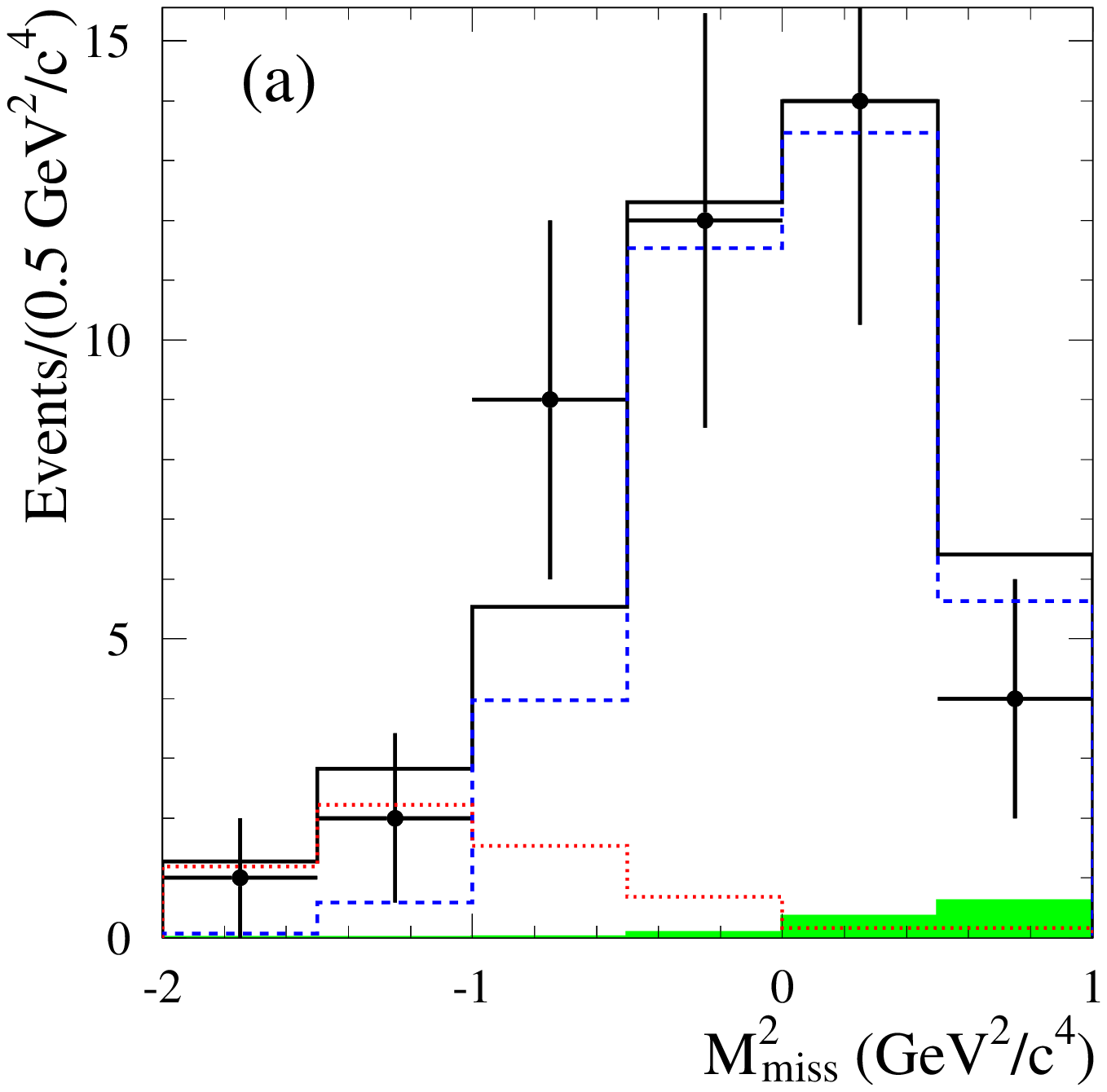}
\includegraphics[width=.32\textwidth]{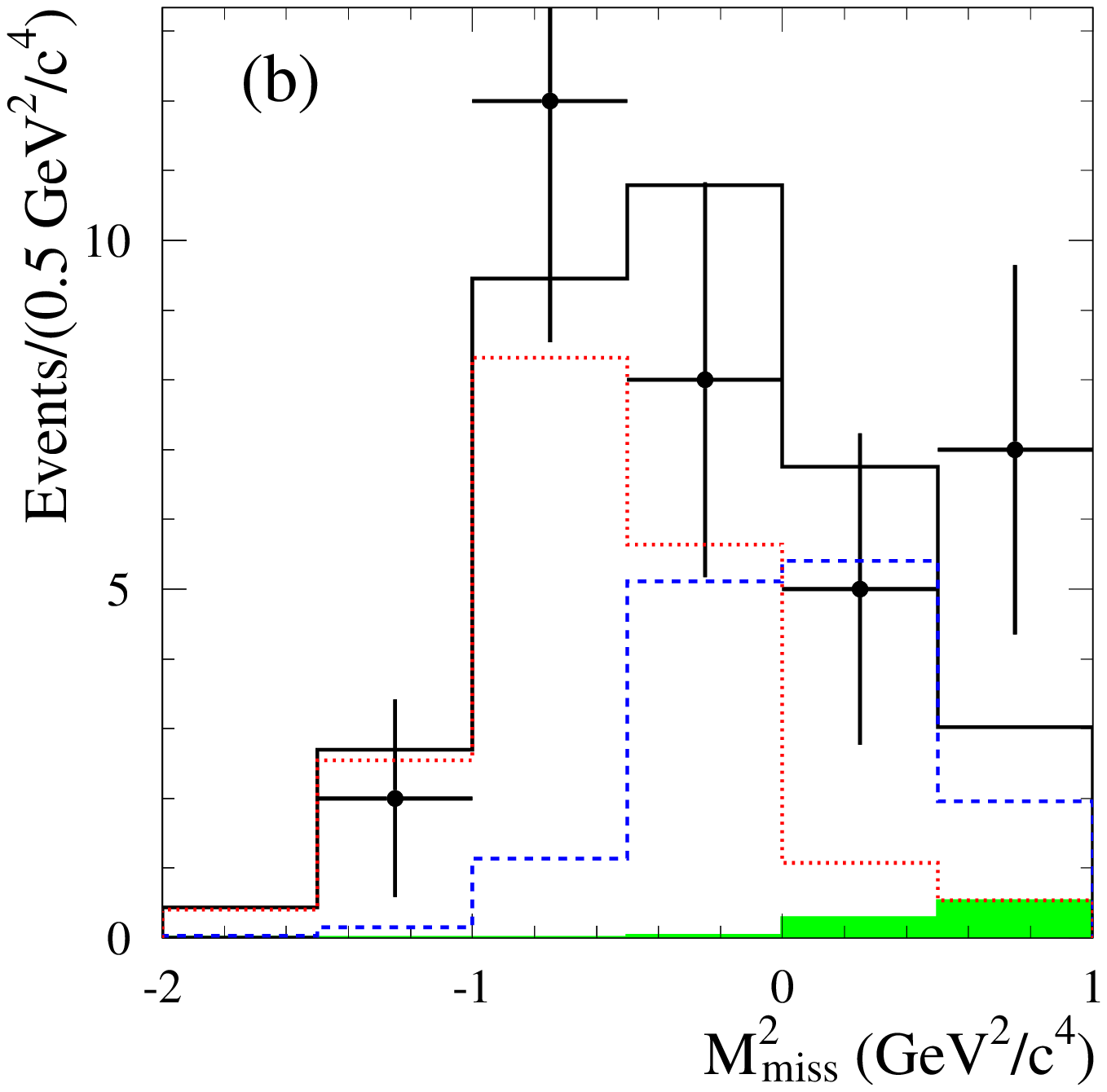}
\includegraphics[width=.32\textwidth]{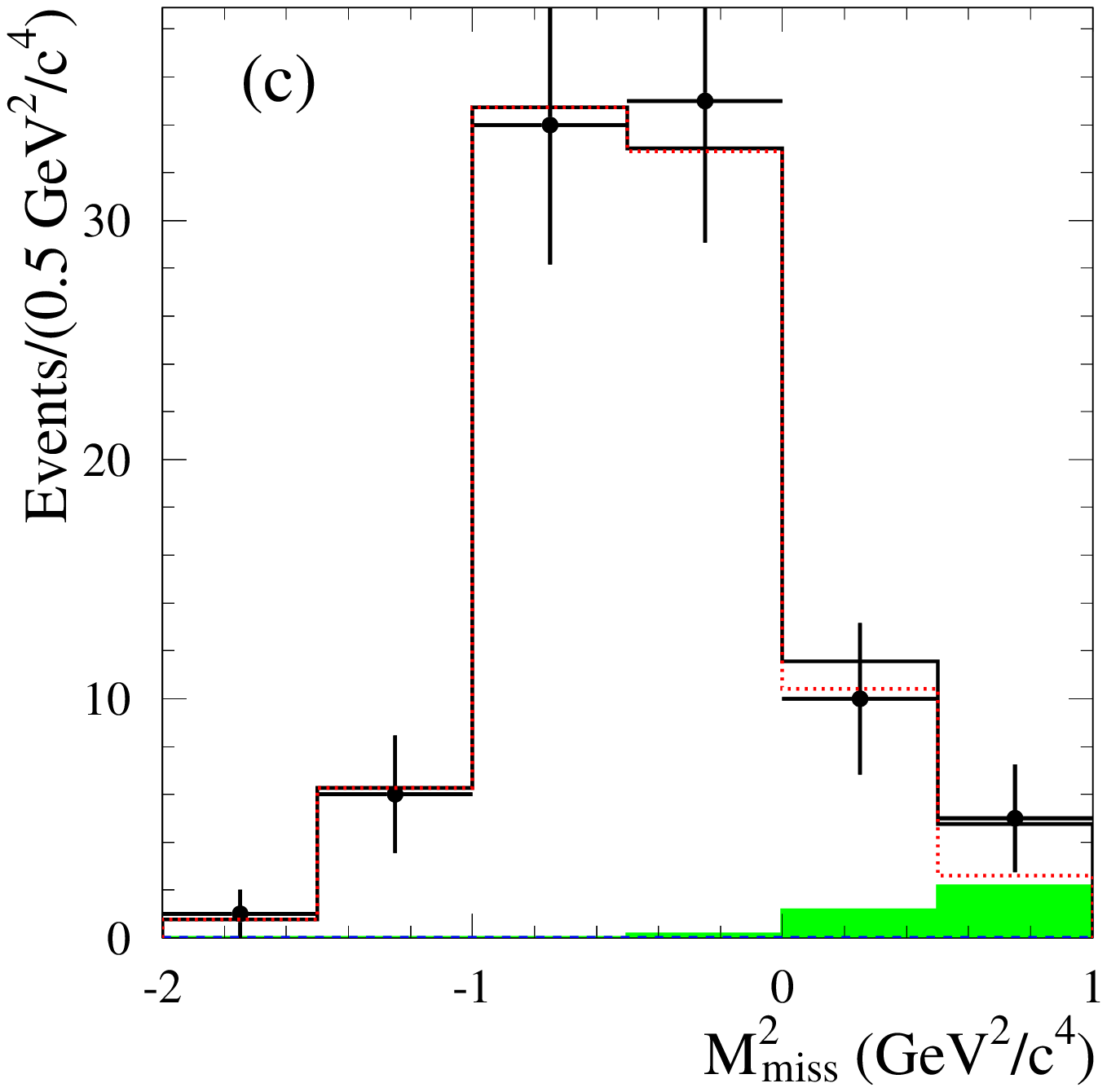}
\caption{(Color online) The $M_{\rm{miss}}^2$ distributions for data 
(points with error
bars) in three $M_{K^+K^-}$ intervals: (a) 5.5--6.0 GeV/$c^2$,
(b) 6.5--7.0 GeV/$c^2$, and (c) 7.5--8.0 GeV/$c^2$. The solid histogram is
the result of the fit described in the text. The dashed blue, dotted red, and 
shaded histograms show the contributions of signal, muon background, and 
ISR + two-photon background, respectively.
\label{fig9}}
\end{figure*}
\begin{table*}
\caption{The number of selected $K^+K^-$ candidates ($N_{\rm data}$),
number of signal events ($N_{\rm sig}$), and
estimated numbers of background events from 
$e^+e^-\to\mu^+\mu^-\gamma$ ($N_{\mu\mu\gamma}$),
from the two-photon process $e^+e^-\to e^+e^-K^+K^-$
($N_{\gamma\gamma}$), and from ISR processes with extra neutral
particle(s) such a $e^+e^-\to K^+K^-\pi^0\gamma$ and $K^+K^-2\pi^0\gamma$
($N_{\rm ISR}$). 
In the last column, $N_{\psi,\chi}$ refers to the background
from $J/\psi\to K^+K^-$ events for $3.0< M_{K^+K^-} <3.2$ GeV/$c^2$, 
from $\psi(2S)\to K^+K^-$ events for $3.6 <M_{K^+K^-}<3.8$ GeV/$c^2$,
and from $\psi(2S)\to \chi_{cJ}\gamma\to K^+K^-\gamma$ events  
for $3.2< M_{K^+K^-} <3.4$ and $3.4 < M_{K^+K^-}< 3.6$ GeV/$c^2$
(see Fig.~\ref{fig14}). 
Events with $M_{K^+K^-}>5.5$ GeV/$c^2$ are selected with the 
looser condition $-2<M_{\rm{miss}}^2<1$ GeV$^2/c^4$.
For $N_{\rm sig}$, the first uncertainty is statistical and the second is
systematic. For the numbers of background events, the combined uncertainty
is quoted.
\label{tab3}}
\begin{ruledtabular}
\begin{tabular}{ccccccc}
$M_{K^+K^-}$ (GeV/$c^2$) & $N_{\rm data}$ & $N_{\rm sig}$ &
$N_{\mu\mu\gamma}$ & $N_{\gamma\gamma}$ & $N_{\rm ISR}$ & $N_{\psi,\chi}$ \\
\hline
\\[-2.1ex]
2.6--2.7 &  76 & $ 75\pm  9\pm 2    $  & $<0.1$       & $<2$         & $0.6\pm 0.5$ & --         \\
2.7--2.8 & 123 & $121\pm 11\pm 2    $  & $<0.1$       & $<2$         & $1.6\pm 1.0$ & --         \\
2.8--2.9 & 160 & $157\pm 13\pm 2    $  & $<0.1$       & $2.6\pm 1.9$ & $0.9\pm 0.7$ & --  \\
2.9--3.0 & 157 & $152\pm 13\pm 2    $  & $<0.1$       & $3.7\pm 2.1$ & $1.3\pm 0.9$ & -- \\ 
3.0--3.2 & 614 & $297\pm 22\pm 3    $  & $<0.1$       & $7.3\pm 2.8$ & $2.3\pm 1.6$ & $307.1\pm 21.3$ \\
3.2--3.4 & 290 & $279\pm 17\pm 2    $  & $<0.1$       & $5.1\pm 2.1$ & $1.8\pm 1.3$ & $4.6\pm 1.6$ \\
3.4--3.6 & 237 & $194\pm 16\pm 2    $  & $<0.1$       & $6.1\pm 1.8$ & $3.1\pm 2.0$ & $33.7\pm 13.8$ \\
3.6--3.8 & 212 & $162\pm 16\pm 1    $  & $<0.1$       & $3.2\pm 0.9$ & $1.5\pm 1.0$ & $45.8\pm 11.0$ \\
3.8--4.0 & 156 & $152\pm 13\pm 1    $  & $<0.1$       & $2.6\pm 0.6$ & $1.4\pm 1.0$ & -- \\
4.0--4.2 & 108 & $105\pm 11\pm 1    $  & $<0.1$       & $2.8\pm 0.5$ & $0.3\pm 0.4$ & -- \\
4.2--4.4 &  84 & $81\pm 9\pm 1      $  & $0.2\pm 0.1$ & $1.2\pm 0.2$ & $1.7\pm 1.0$ & -- \\
4.4--4.6 &  47 & $44.7\pm 6.9\pm 0.6$  & $0.1\pm 0.1$ & $1.2\pm 0.2$ & $1.0\pm 0.7$ & -- \\
4.6--4.8 &  43 & $41.2\pm 6.6\pm 0.3$  & $0.1\pm 0.1$ & $1.5\pm 0.3$ & $0.2\pm 0.3$ & -- \\
4.8--5.0 &  38 & $36.2\pm 6.2\pm 0.5$  & $0.5\pm 0.3$ & $0.8\pm 0.2$ & $0.5\pm 0.4$ & -- \\
5.0--5.2 &  28 & $26.8\pm 5.3\pm 0.3$  & $0.2\pm 0.1$ & $0.6\pm 0.1$ & $0.4\pm 0.4$ & -- \\
5.2--5.5 &  47 & $45.2\pm 6.9\pm 0.6$  & $0.9\pm 0.5$ & $0.6\pm 0.2$ & $0.3\pm 0.3$ & -- \\
\hline
5.5--6.0 &  42 & $35.3\pm 6.7 \pm 0.7$ & $6.0\pm3.7$  & $0.4\pm0.3$  & $0.7\pm0.5$  &  -- \\
6.0--6.5 &  25 & $10.9\pm 4.6 \pm 1.2$ & $11.4\pm4.3$ & $0.3\pm0.2$  & $2.0\pm1.1$  &  -- \\
6.5--7.0 &  34 & $13.8\pm 5.4 \pm 0.7$ & $18.5\pm5.6$ & $<0.3$       & $0.8\pm0.6$  &  -- \\
7.0--7.5 &  44 & $ 7.5\pm 5.3 \pm 1.9$ & $33.3\pm6.9$ & $<0.5$       & $3.4\pm1.8$  &  -- \\
7.5--8.0 &  91 & $ 0.0\pm 7.0 \pm 2.0$ & $87.6\pm10.5$& $<0.5$       & $3.5\pm1.9$  &  -- \\
\end{tabular}
\end{ruledtabular}
\end{table*}
To be selected and thus to represent background for
this analysis, both final-state charged tracks in 
$e^+e^-\to e^+e^-\gamma$, $\mu^+\mu^-\gamma$, and 
$\pi^+\pi^-\gamma$ events must be misidentified as kaons and, and the
missing-mass squared must be poorly determined.

The probability to misidentify a pion as a kaon has been measured 
as a function of charge, momentum, and polar angle using a control sample 
of pions from $K_S\to \pi^+\pi^-$
decays. Using the measured misidentification probabilities, we calculate  
weights for simulated $e^+e^-\to \pi^+\pi^-\gamma$ events 
(see Sec.~\ref{detector}) to be identified as $K^+K^-\gamma$ events, and
estimate a $\pi^+\pi^-\gamma$ background rate 
relative to the signal $K^+K^-\gamma$ rate ranging from
$5\times10^{-5}$ at 3 GeV/$c^2$ to about $5\times10^{-3}$ at 7.5 GeV/$c^2$.

A similar approach is used to estimate the $e^+e^-\to e^+e^-\gamma$ 
background. The electron misidentification rate has been measured using 
$e^+e^-\to e^+e^-\gamma$
events with the photon detected at large angles. From MC simulation we 
estimate the electron contamination to be 
at most 0.5\%. The PID requirements suppress $e^+e^-\to e^+e^-\gamma$ events
by a factor of about $10^{8}$. 
We have verified this suppression by analyzing a sample of LA  
$K^+K^-\gamma$ candidates with the photon detected in the EMC. 
In this data sample, surviving $e^+e^-\to e^+e^-\gamma$ events 
can be identified by requiring a small opening angle between the photon
direction and that of one of the charged-particle tracks. 

In the subsequent analysis, we disregard possible backgrounds
from $e^+e^-\gamma$ and $\pi^+\pi^-\gamma$ events 
since their contributions are expected to be negligible.

The $e^+e^-\to \mu^+\mu^-\gamma$ background is non-negligible for large values
of $M_{K^+K^-}$.
For $M_{K^+K^-}> 5.5$ GeV/$c^2$, we estimate the numbers of signal and 
background events
in each of the five mass intervals listed in Table~\ref{tab3} by fitting the 
$M_{\rm{miss}}^2$ distributions in the range $[-2,+1]$ GeV$^2/c^4$, as shown 
in Fig.~\ref{fig9}, using three components: signal events,
the $\mu^+\mu^-\gamma$ background, and the ISR + two-photon background.
The $M_{\rm{miss}}^2$ interval is extended to negative values to increase
the sensitivity to $e^+e^-\to \mu^+\mu^-\gamma$ background and thus to 
better determine its contribution.
The distribution for signal 
events is taken from simulation and is centered at zero.  The distribution
for the $\mu^+\mu^-\gamma$ background is obtained using data events
with at least one identified muon, and is
shifted to negative $M_{\rm{miss}}^2$ values because of the muon-kaon mass
difference. We also include the small contributions from  ISR and two-photon 
events estimated as described below in Secs.~\ref{ISRbkg} and \ref{ggbkg}.
The fitted parameters are the numbers of signal ($N_{\rm sig}$) 
and muon-background ($N_{\mu\mu\gamma}$) events.

The results of the fits are listed in the last five rows of 
Table~\ref{tab3} and are shown in Fig.~\ref{fig9} for three representative 
intervals of $M_{K^+K^-}$. The first uncertainty in $N_{\rm sig}$ 
is statistical, while the second, systematic, uncertainty, accounts for the
uncertainty
in the numbers of ISR and two-photon background events.
The $N_{\rm sig}$ results in Table~\ref{tab3}
for $M_{K^+K^-}>5.5$ GeV/$c^2$
are obtained with the condition $-2<M_{\rm{miss}}^2<1$ GeV$^2$/$c^4$. They 
can be scaled into our standard selection $|M_{\rm{miss}}^2|<1$ GeV$^2$/$c^4$ 
by multiplying the results in the 5.5--6.5 GeV/$c^2$ 
and 6.5--7.5 GeV/$c^2$ mass ranges 
by 0.98 and 0.99, respectively. For $M_{K^+K^-}>7.5$
GeV/$c^2$ the scaling factor is consistent with 1.0.
The scale factors are determined using simulated signal events.

Below 5.5 GeV/$c^2$, where the muon background is small, we adopt a simpler 
approach and estimate the number of $\mu^+\mu^-\gamma$ background events 
in each mass interval
using the number of selected events $N_{1\mu}$ with at least one charged track
identified as a muon. The number of background events is estimated as
\begin{equation}
N_{\mu\mu\gamma}=C_\mu (N_{1\mu}-k_{1\mu}N_{\rm data}),
\label{mubkg}
\end{equation}
where $C_\mu$, evaluated as described below, is the ratio of the number of 
selected $\mu^+\mu^-\gamma$ events with no identified muon
to the number of events with at least one identified muon, $k_{1\mu}$ 
is the fraction of selected $K^+K^-\gamma$ events with at least one identified
muon, and $N_{\rm data}$ is the number of events in the respective $M_{K^+K^-}$
interval.
The value of $k_{1\mu}$ is taken from simulated signal events and varies 
from 0.006 at $M_{K^+K^-}=2.6$ GeV$/c^2$ to 0.01 at
$M_{K^+K^-}=5.5$ GeV/$c^2$.
\begin{figure*}
\includegraphics[width=.32\textwidth]{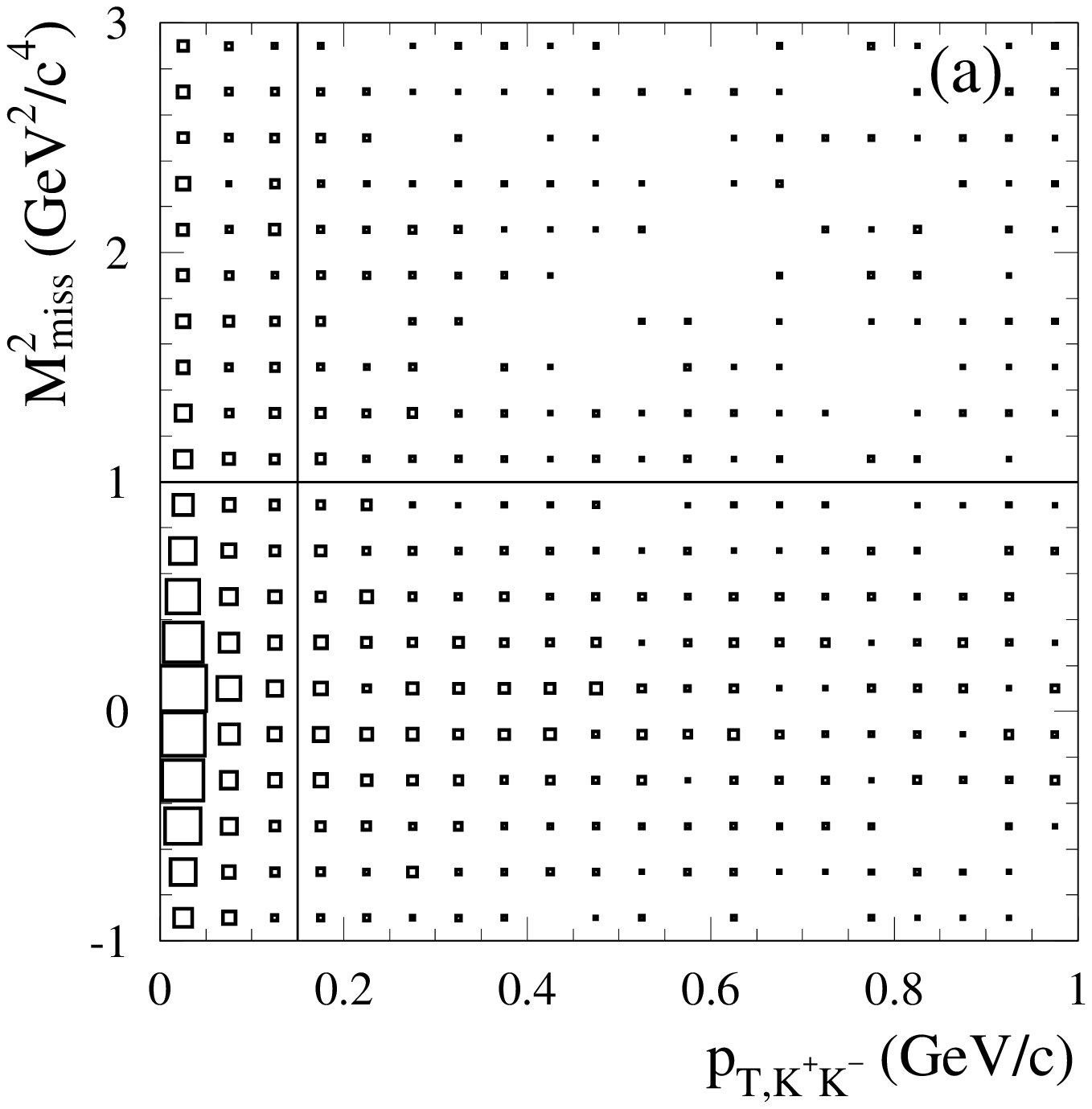}
\includegraphics[width=.32\textwidth]{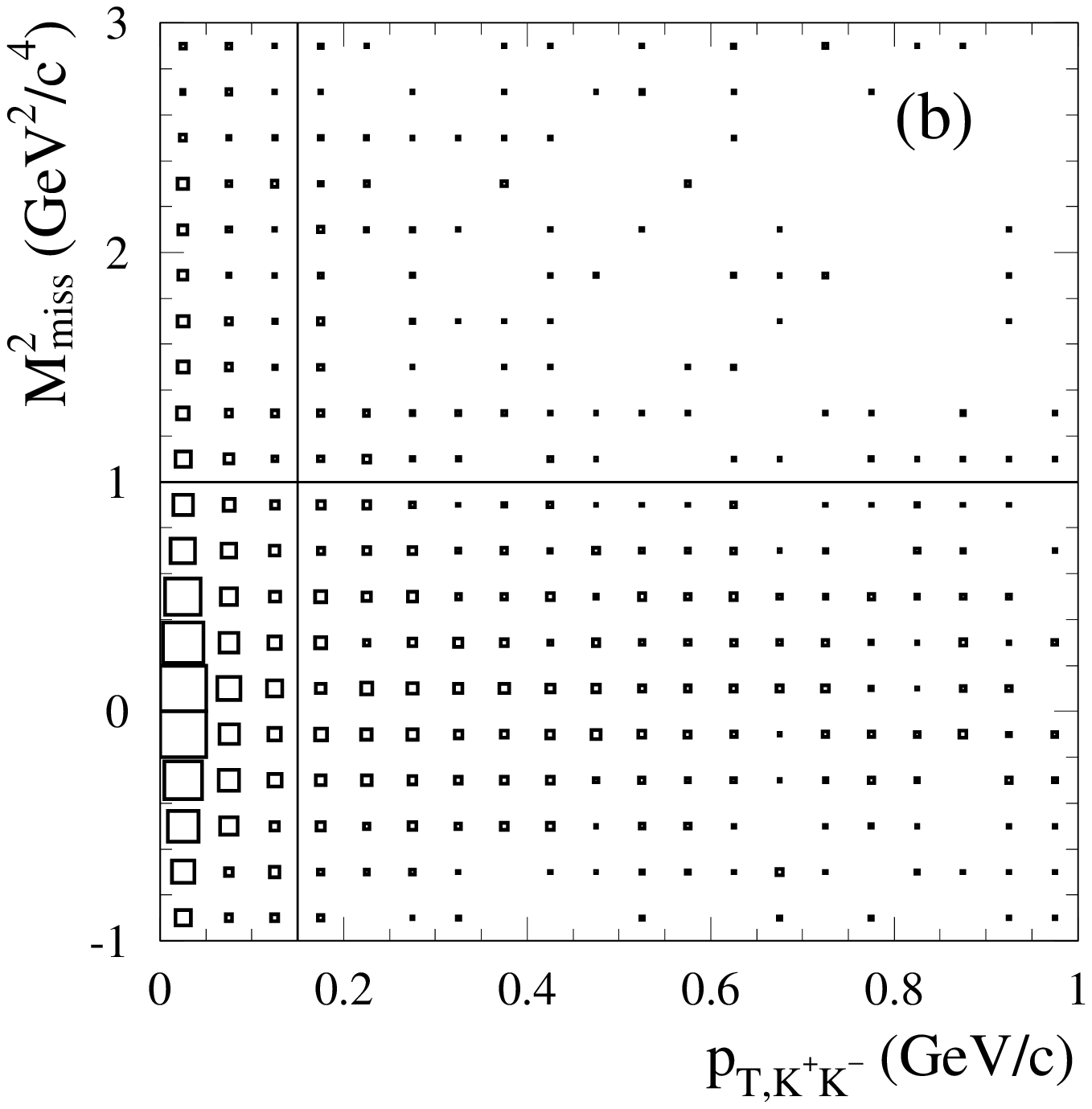}
\includegraphics[width=.32\textwidth]{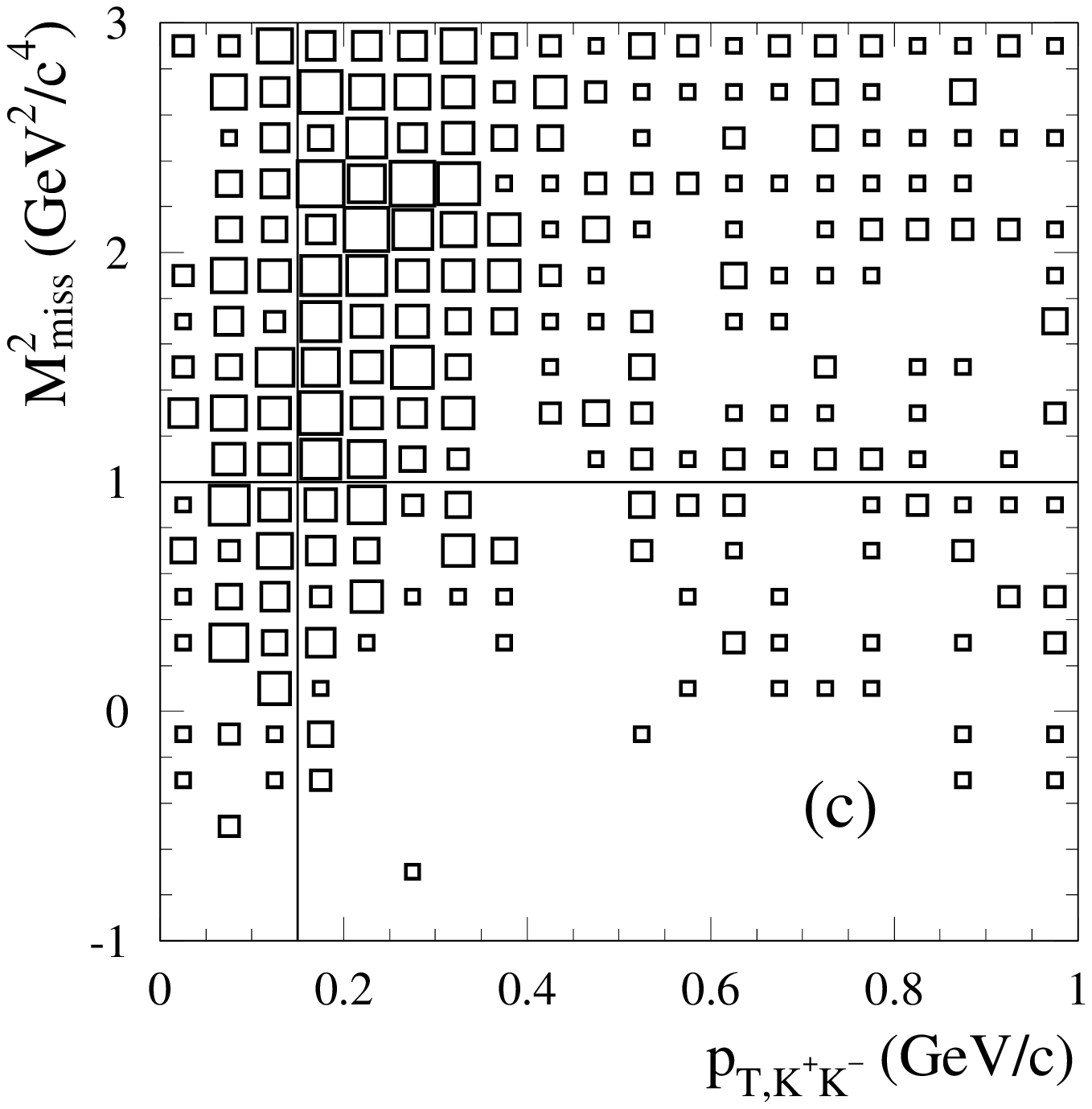}
\caption{The distributions of $M_{\rm{miss}}^2$ versus $p_{{\rm T},K^+K^-}$
for (a) data events, (b) simulated signal events, and (c) simulated ISR 
background events.
Events in the $J/\psi$ and $\chi_{c0}$ mass regions, 
$3.05 < M_{K^+K^-} < 3.15$ GeV/$c^2$ and 
$3.38<M_{K^+K^-}<3.46$ GeV/$c^2$, are excluded from the distributions;
regions near the $\chi_{c2}$ and $\psi(2S)$
are not excluded, since their signal 
content is quite small. The lines indicate the boundaries
of the signal region (bottom left rectangle) and the sideband region
(top right rectangle).
\label{fig13}}
\end{figure*}

Very few simulated $\mu^+\mu^-\gamma$ events have both tracks identified as
kaons and neither identified as a muon, so $C_\mu$ is studied as a function 
of $M_{K^+K^-}$ using the probability for an individual muon to be
identified as both a kaon and a muon, assuming the probabilities for
the two tracks to be independent.  
We find that $C_\mu$ does not exhibit a significant dependence on mass 
within the range of our measurements, 2.6--8.0 GeV/$c^2$.
Therefore, $C_\mu$ used in Eq.~(\ref{mubkg}) is estimated
from the fitted numbers of $\mu^+\mu^-\gamma$ events above 5.5 GeV/$c^2$.
We find $C_\mu=0.14\pm 0.01\pm 0.08$, where the first 
uncertainty is from the fits and the second accounts for
the full range of values in different mass 
intervals in data and simulation (for purposes of information, the
MC result is $C_\mu=0.11$). The resulting estimated 
numbers of $\mu^+\mu^-\gamma$ background events are listed in 
Table~\ref{tab3}. For masses below 4.2 GeV/$c^2$,  
$(N_{1\mu}-k_{1\mu}N_{\rm data})$ is consistent with zero, and
we take 0.1 as both an upper limit and uncertainty.
\begin{figure*}
\includegraphics[width=.4\textwidth]{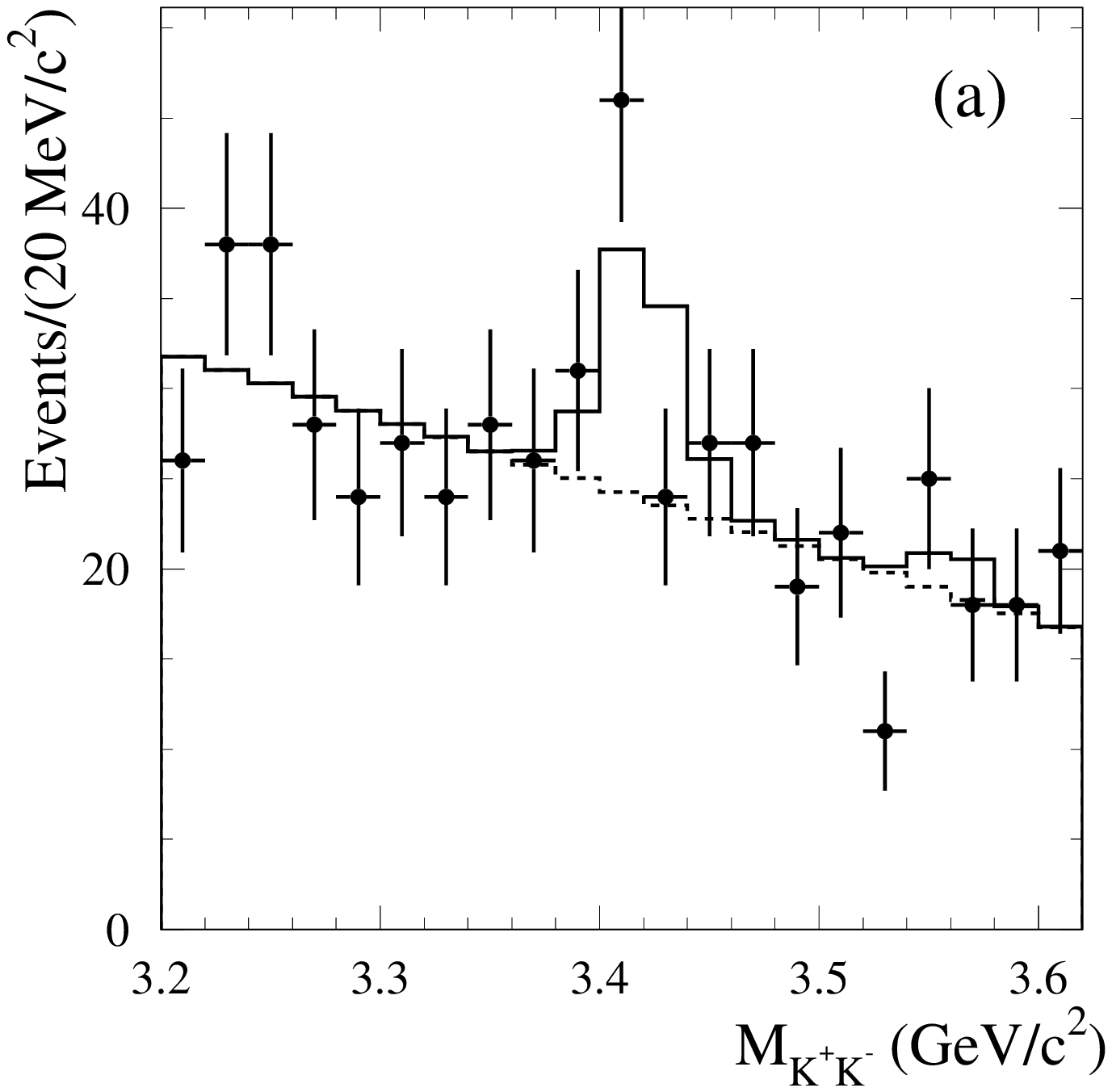}
\includegraphics[width=.4\textwidth]{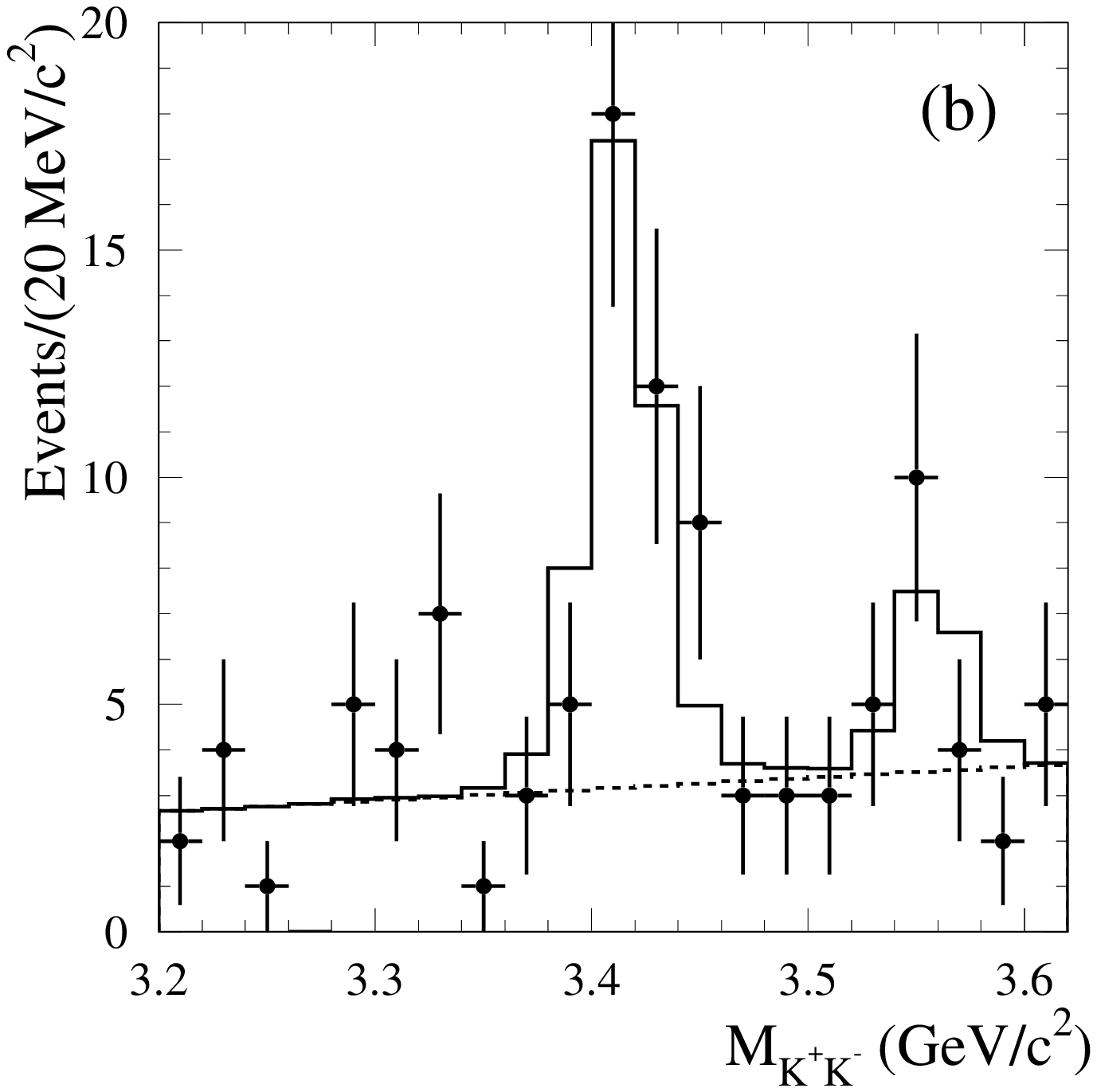}
\caption{The $M_{K^+K^-}$ spectra for data (points with error bars) in 
the vicinity of the $\chi_{c0}$ and $\chi_{c2}$ resonances for the 
$M_{\rm{miss}}^2$--$p_{{\rm T},K^+K^-}$
(a) signal region and (b) sideband region. The solid histograms result 
from the fits described in the text. The dashed histograms represent 
the nonresonant contributions.
\label{fig14}}
\end{figure*}

\subsection{ \boldmath Multibody ISR background\label{ISRbkg}}
Background ISR events containing a $K^+K^-$ pair and one or more
$\pi^0$ and/or $\eta$ mesons are distinguishable by their nonzero 
values of $M_{\rm{miss}}^2$ and $p_{{\rm T},K^+K^-}$, but some events 
with a small number of neutral particles still can enter the selected
data sample. Figure~\ref{fig13}(a) shows the two-dimensional distribution
of $M_{\rm{miss}}^2$ versus $p_{{\rm T},K^+K^-}$ for data events before 
the requirements on these two variables, indicated by the lines, are applied.
The bottom left rectangle is the signal region.
The same distribution for simulated signal events is shown in
Fig.~\ref{fig13}(b), and is similar in structure except for a deficit in the 
upper right rectangle, which we take as a sideband region. The distribution 
for ISR background events
produced by JETSET is shown in Fig.~\ref{fig13}(c). It should be noted that
most (98\%) simulated background events in the signal region are 
from the process $e^+e^-\to K^+K^-\pi^0\gamma$, while the fraction 
in the sideband region is about 80\%.

The number of data events in the sideband region $N_2$ is used to
estimate the ISR background in the signal region using
\begin{equation}
N_{\rm ISR}=\frac{N_2-\beta_{\rm sig}N_1}{\beta_{\rm bkg}-\beta_{\rm sig}},
\label{bkgisr}
\end{equation}
where $N_1$ is the number of data events in the signal region, and
$\beta_{\rm sig}$ and $\beta_{\rm bkg}$ are the $N_2/N_1$ ratios from signal
and background simulation, respectively. The coefficient $\beta_{\rm sig}$ 
increases linearly from  $0.046\pm0.005$ at
$M_{K^+K^-}=2.6$ GeV/$c^2$ to $0.074\pm0.005$ at 8.0 GeV/$c^2$,
where the uncertainty is statistical,
whereas $\beta_{\rm bkg}=7.6\pm1.0\pm4.0$ is independent of $M_{K^+K^-}$.
The first uncertainty in $\beta_{\rm bkg}$ is statistical, and the second is 
systematic. The latter takes into account  possible differences
between data and simulation in the background composition, and in
the kinematic distributions of $e^+e^-\to K^+K^-\pi^0\gamma$ events.

The regions 3.0--3.2 and 3.6--3.8 GeV/$c^2$ contain resonant contributions from
the decays $J/\psi\to K^+K^-$ and $\psi(2S)\to K^+K^-$, respectively. 
The resonant and nonresonant contributions are determined by the fits
described in Sec.~\ref{psi}, and such fits are also applied to the sideband
regions.  The resulting numbers of nonresonant events, $N_1$ and $N_2$,
are used to estimate the ISR background.

Similarly, the $M_{K^+K^-}$ regions 3.2--3.4 and 3.4--3.6 GeV/$c^2$ 
contain $\chi_{c0}$ and
$\chi_{c2}$ decays, as seen in Fig.~\ref{fig14}. The $\chi_{cJ}$ states 
are produced in the reaction $e^+e^-\to \psi(2S)\gamma$,  followed by
$\psi(2S)\to \chi_{cJ}\gamma$.   
A similar set of fits is used to determine $N_1$, $N_2$, and the 
background contribution from the $\chi_{cJ}$ states, and the fit results are 
shown in Fig.~\ref{fig14}.  
The estimated numbers of ISR background events are listed in 
Table~\ref{tab3} along with the fitted numbers of 
$\psi$ and $\chi_{cJ}$ decays in the relevant mass intervals. 

\subsection{ \boldmath Two-photon background\label{ggbkg}}
\begin{figure}
\includegraphics[width=.4\textwidth]{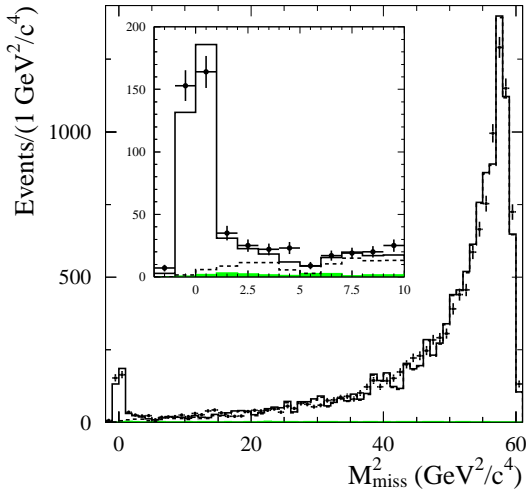}
\caption{The $M_{\rm{miss}}^2$ distribution for data (points with error
bars) with $2.8<M_{K^+K^-}<3.0$ GeV/$c^2$ selected with all the criteria 
described in Sec.~\ref{sel} except for the requirement $|M_{\rm{miss}}^2|<1$
GeV$^2$/$c^4$.
The solid histogram is a sum of signal and background distributions obtained
from MC simulation.
The dashed histogram shows the distribution for two-photon events, and 
the shaded (almost invisible) histogram shows the small contribution
of all other background processes. The 
inset shows an enlarged view of the region 
$-2<M_{\rm{miss}}^2<10$ GeV$^2$/$c^4$.
Above 10 GeV$^2$/$c^4$ the solid and dashed histograms are indistinguishable.
\label{fig11}}
\end{figure}
\begin{figure}
\includegraphics[width=.4\textwidth]{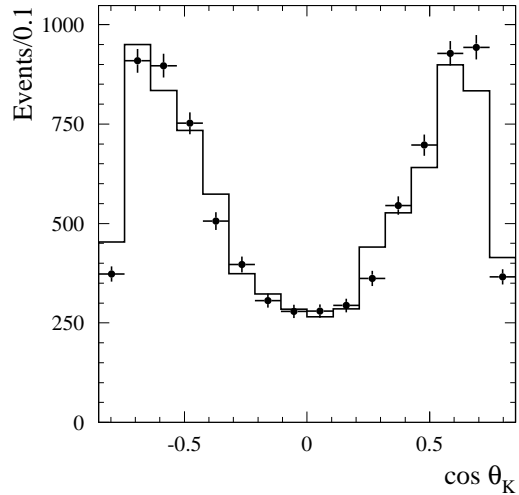}
\caption{The $\cos{\theta_K}$ distribution for data (points with error bars)
and reweighted simulated events (histogram) with $M_{\rm{miss}}^2>20$
GeV$^2/c^4$ from the $K^+K^-$ mass range 2.8--3.0 GeV/$c^2$. 
\label{fig12}}
\end{figure}
Two-photon events corresponding to the process
$e^+e^-\to e^+e^-\gamma^\ast\gamma^\ast\to e^+e^-K^+K^-$
are distinguished by their larger values of $M_{\rm{miss}}^2$.
Figure~\ref{fig11} shows the $M_{\rm{miss}}^2$ distribution for data events 
in the range $2.8<M_{K^+K^-}<3.0$ GeV/$c^2$ that satisfy all the criteria 
in Sec.~\ref{sel} except for that on $M_{\rm{miss}}^2$. The two-photon events,
which dominate the large $|M_{\rm{miss}}^2|$ region, are generally seen to be
well separated from signal events but to nonetheless have a tail that extends
into the signal region $|M_{\rm{miss}}^2|<1$ GeV$^2/c^4$.
The exact shape of this tail depends on the unknown kaon angular 
distribution.  Therefore, we reweight our simulation (generated with a
uniform distribution) to reproduce the $\cos{\theta_K}$ distribution 
observed in the data in each 
$M_{K^+K^-}$ interval; here $\theta_K$ is the angle between the $K^{+}$
momentum in the $K^+K^-$ rest frame and the $e^-$ beam direction in 
the $e^+e^-$ 
c.m.~frame. The data and reweighted simulated $\cos{\theta_K}$ distributions 
are compared
in Fig.~\ref{fig12}. The simulated $M_{\rm{miss}}^2$ distribution is
shown in Fig.~\ref{fig11}, where it is seen to reproduce the data well.

The two-photon background in each $M_{K^+K^-}$ interval is estimated from the number of 
data events with $M_{\rm{miss}}^2 > d$ and a scale factor from the simulation.
The $M_{\rm{miss}}^2$ distribution changes with $M_{K^+K^-}$, and the value of
$d$ is 20 GeV$^2/c^4$ for $M_{K^+K^-}<4.4$ GeV/$c^2$, 10 GeV$^2/c^4$ for
$M_{K^+K^-}>6.5$ GeV/$c^2$, and varies linearly in-between.  The scale factor
ranges from $10^{-4}$ in the 2.6--2.7 GeV/$c^2$ interval to about $10^{-2}$
in the 7.0--7.5 GeV/$c^2$ interval. However, the number of two-photon events 
decreases with increasing $M_{K^+K^-}$. The estimated 
background event contributions are listed in Table~\ref{tab3}.

The numbers of signal events obtained after background subtraction are 
listed in Table~\ref{tab3}. 
The first uncertainty in $N_{\rm sig}$ is statistical and the second is 
systematic. The systematic term  accounts for the uncertainties in the numbers
of $e^+e^- \to \mu^+\mu^-\gamma$ and two-photon background 
events, and the uncertainties in the coefficients $\beta_{\rm sig}$ and 
$\beta_{\rm bkg}$ in the ISR background subtraction procedure.

\section{Detection efficiency\label{deteff}}
\begin{figure}
\includegraphics[width=.4\textwidth]{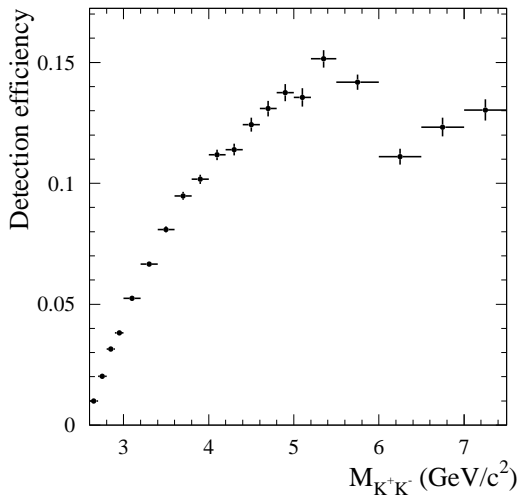}
\caption{The $K^+K^-$ mass dependence of the detection
efficiency for simulated $e^+e^-\to K^+K^-\gamma$ events. 
\label{fig15}}
\end{figure}
The detection efficiency, $\varepsilon_{\rm MC}$, determined using MC 
simulation, is shown in Fig.~\ref{fig15} as a function of $M_{K^+K^-}$. 
The nonmonotonic behavior observed for $M_{K^+K^-}>5.5$ GeV/$c^2$ is 
introduced by filters designed to reduce background
before the event-reconstruction stage.

Corrections are applied to $\varepsilon_{\rm MC}$ to account for
data-MC simulation differences in detector response
\begin{equation}
\varepsilon=\varepsilon_{\rm MC}\prod_{i=1}^{4} (1+\delta_i), 
\label{eq_eff_cor}
\end{equation}
where the $\delta_i$ terms are the efficiency corrections listed in 
Table~\ref{tab_ef_cor}. 
\begin{table*}
\caption{Values of the efficiency corrections, $\delta_i$.
The three values in the rows ``PID'' and ``total'' correspond to 
$M_{K^+K^-}=2.6$, 6.0, and 7.5 GeV/$c^2$, respectively.
\label{tab_ef_cor}}
\begin{ruledtabular}
\begin{tabular}{lc}
Effect               &$\delta_i$ (\%) \\
\hline\\[-2.1ex]
Trigger              & $-1.0\pm 0.5$ \\  
PID                  & $-2.0\pm 0.4$/$-4.0\pm 0.5$/$-10.0\pm 1.7$ \\
Track reconstruction & $ 0.0\pm 1.6$  \\
Requirements on $p_{{\rm T},K^+K^-}$ and $M_{\rm{miss}}^2$ & $ 0.0\pm 1.5$ \\
\hline
Total                & $-3.0\pm2.3$/$-5.0\pm2.3$/$-11.0\pm2.8$ \\
\end{tabular}
\end{ruledtabular}
\end{table*}
The difference between data and simulation in trigger efficiency is 
studied using the overlap of the samples of events satisfying two independent
sets of trigger criteria based on signals from the EMC and DCH.
The correction for trigger inefficiency is found to be 
$(-1.0\pm 0.5)\%$.  

The track-reconstruction efficiency for charged kaons has been studied in 
events with 
similar topology~\cite{KKbabar} and we use the correction derived 
therein.  The charged-kaon identification efficiency is studied as a function
of the track momentum and polar angle using a control sample of kaons from the 
decay chain 
$D^{\ast +} \rightarrow \pi^{+} D^{0}, D^{0} \rightarrow K^{-} \pi^{+}$.
The ratio of the efficiencies is then used to reweight simulated signal
events, resulting in an overall correction that
varies slowly from from $-2\%$ at 2.6 GeV/$c^2$ to $-4\%$
at 6 GeV/$c^2$, and then falls to about $-10\%$ at 7.5 GeV/$c^2$.
The statistical uncertainty in the correction term defines 
the systematic uncertainty in this correction.

The remaining criteria are based on $M_{\rm{miss}}^2$ and $p_{{\rm T},K^+K^-}$,
which we believe to be well simulated. The track momentum and angular 
resolutions have been studied in, e.g., Ref.~\cite{ppbar1}
for $e^+e^-\to \mu^+\mu^-\gamma$ events with a detected photon.  Based 
on this and similar variables in our previous ISR studies, we make no 
correction, and assign a conservative systematic uncertainty of 1.5\%
to cover these remaining factors.
The corrections to the detection efficiency are listed in 
Table~\ref{tab_ef_cor}.

\section{ \boldmath The $e^+e^-\to K^+K^-$ cross section and 
the charged-kaon form factor}\label{crosssec}
\begin{figure}
\includegraphics[width=.4\textwidth]{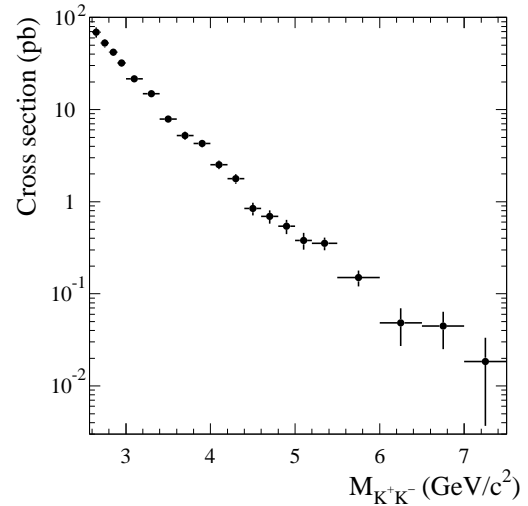}
\caption{The $e^+e^-\to K^+K^-$ cross section measured
in this work.
\label{figcs}}
\end{figure}
\begin{figure}
\includegraphics[width=.4\textwidth]{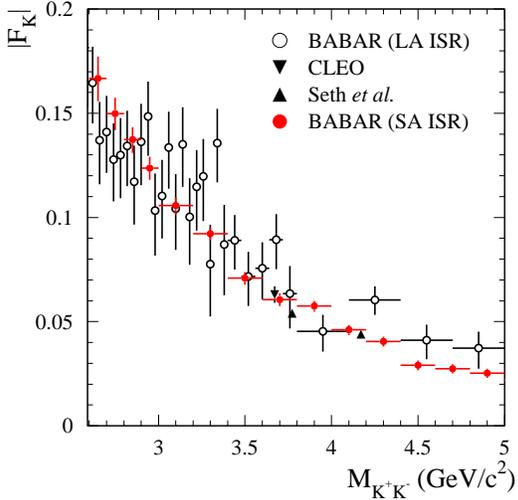}
\caption{The charged-kaon electromagnetic form factor measured
in this analysis [BABAR (SA ISR)] in comparison with previous measurements:
CLEO~\cite{CLEO}, Seth {\it et al}.~\cite{NU}, and BABAR (LA ISR)~\cite{KKbabar}
in the mass region 2.6--5.0 GeV/$c^2$. Only statistical uncertainties
are shown.
\label{fig19}}
\end{figure}
\begin{table*}
\caption{The $K^+K^-$ invariant-mass interval ($M_{K^+K^-}$),
number of selected events ($N_{\rm sig}$) after background subtraction,
detection efficiency ($\varepsilon$), 
ISR luminosity ($L$), measured  $e^+e^-\to K^+K^-$ cross section
($\sigma_{K^+K^-}$), and the charged-kaon form factor  
($|F_K|$). For the number of events and cross section, the first uncertainty
is statistical and the second is systematic. For the form factor, we quote 
the combined uncertainty.
For the mass interval 7.5--8.0 GeV/$c^2$, the 90\% CL upper limits for the
cross section and form factor are listed.
\label{sumtab}}
\begin{ruledtabular}
\begin{tabular}{cccccc}
$M_{K^+K^-}$ &$N_{\rm sig}$&$\varepsilon$&$L$&$\sigma_{K^+K^-}$&$|F_K|\times 100$\\
(GeV/$c^2$)& & (\%) & (pb$^{-1}$) & (pb)& \\
\hline\\[-2.1ex]
2.6--2.7 & $ 75\pm  9\pm 2    $ & 0.96 & 113 & $69.3\pm 8.1\pm 3.7     $ & $16.7\pm 1.1       $ \\
2.7--2.8 & $121\pm 11\pm 2    $ & 1.94 & 118 & $52.9\pm 4.9\pm 2.4  	 $ & $15.0\pm 0.8	   $ \\
2.8--2.9 & $157\pm 13\pm 2    $ & 3.03 & 122 & $42.0\pm 3.4\pm 1.6  	 $ & $13.7\pm 0.6	   $ \\
2.9--3.0 & $152\pm 13\pm 2    $ & 3.69 & 127 & $32.2\pm 2.7\pm 1.2  	 $ & $12.4\pm 0.6	   $ \\
3.0--3.2 & $297\pm 22\pm 3    $ & 5.07 & 271 & $21.7\pm 1.6\pm 0.6  	 $ & $10.6\pm 0.4	   $ \\
3.2--3.4 & $279\pm 17\pm 2    $ & 6.43 & 292 & $14.8\pm 0.9\pm 0.4  	 $ & $9.2\pm 0.3	   $ \\
3.4--3.6 & $194\pm 16\pm 2    $ & 7.81 & 313 & $7.92\pm 0.63\pm 0.24   $ & $7.1\pm 0.3	   $ \\
3.6--3.8 & $162\pm 16\pm 1    $ & 9.15 & 336 & $5.26\pm 0.51\pm 0.16   $ & $6.1\pm 0.3	   $ \\
3.8--4.0 & $152\pm 13\pm 1    $ & 9.80 & 361 & $4.30\pm 0.36\pm 0.13   $ & $5.7\pm 0.3	   $ \\
4.0--4.2 & $105\pm 11\pm 1    $ & 10.8 & 386 & $2.52\pm 0.25\pm 0.08   $ & $4.60\pm 0.25	   $ \\
4.2--4.4 & $81\pm 9\pm 1      $ & 11.0 & 413 & $1.79\pm 0.20\pm 0.06   $ & $4.05\pm 0.25	   $ \\
4.4--4.6 & $44.7\pm 6.9\pm 0.6$ & 11.9 & 442 & $0.85\pm 0.13\pm 0.03   $ & $2.91\pm 0.23	   $ \\
4.6--4.8 & $41.2\pm 6.6\pm 0.3$ & 12.5 & 473 & $0.70\pm 0.11\pm 0.03   $ & $2.74\pm 0.23	   $ \\
4.8--5.0 & $36.2\pm 6.2\pm 0.5$ & 13.1 & 507 & $0.55\pm 0.09\pm 0.02   $ & $2.52\pm 0.22	   $ \\
5.0--5.2 & $26.8\pm 5.3\pm 0.3$ & 12.9 & 543 & $0.38\pm 0.08\pm 0.02   $ & $2.19\pm 0.22	   $ \\
5.2--5.5 & $45.2\pm 6.9\pm 0.6$ & 14.4 & 888 & $0.35\pm 0.05\pm 0.01   $ & $2.21\pm 0.18	   $ \\
5.5--6.0 & $34.6\pm 6.6\pm 0.7$ & 13.5 & 1710 &  $0.150\pm 0.029\pm 0.006$ & $1.54\pm 0.15	   $ \\
6.0--6.5 & $10.7\pm 4.5\pm 1.2$ & 10.6 & 2062 &  $0.049\pm 0.021\pm 0.006$ & $0.95\pm 0.22	   $ \\
6.5--7.0 & $13.6\pm 5.3\pm 0.7$ & 11.6 & 2523 &  $0.047\pm 0.018\pm 0.004$ & $1.00\pm 0.20	   $ \\
7.0--7.5 & $7.4\pm  5.3\pm 1.9$ & 11.9 & 3144 &  $0.020\pm 0.014\pm 0.004$ & $0.70_{-0.36}^{+0.23} $ \\
7.5--8.0 & $0.0\pm  6.9\pm 2.0$ & 9.36 & 4015 &  $<0.024                 $ & $<0.9		   $ \\	      						      	    				
\end{tabular} 	    				
\end{ruledtabular}
\end{table*}   	    				
The $e^+e^-\to K^+K^-$ cross section in each $K^+K^-$ mass interval $i$ is 
calculated as 
\begin{equation}
\sigma_{K^+K^-,i}=\frac{N_{{\rm sig},i}}{\varepsilon_i L_i}.
\end{equation}
The number of selected
events ($N_{{\rm sig},i}$) for each $K^+K^-$ mass interval after 
background subtraction is listed in Table~\ref{sumtab}.
The $N_{\rm sig}$ values for $M_{K^+K^-}>5.5$ GeV/$c^2$ differ
from the corresponding values in Table~\ref{tab3}. They are
corrected to correspond to the nominal selection 
$|M_{\rm{miss}}^2|<1$ GeV$^2$/$c^4$ as described in Sec.~\ref{bkg_mumu}.
The first uncertainty in $N_{\rm sig}$ is statistical; the second is
systematic due to background subtraction.
The value of the ISR luminosity $L_i$ is obtained by integrating
$W(s,x)$ from Refs.~\cite{radf1,radf2} over mass interval $i$ and is 
listed in Table~\ref{sumtab}. The formulas from Refs.~\cite{radf1,radf2}
include higher-order radiative corrections. However, we do not include 
in $W(s,x)$ corrections for leptonic and hadronic vacuum polarization
in the photon propagator. Cross sections obtained in this way
are referred to as ``dressed".

The values obtained for the $e^+e^-\to K^+K^-$ cross section are listed 
in Table~\ref{sumtab}. For the mass intervals
3.0--3.2 GeV/$c^2$ and 3.6--3.8 GeV/$c^2$ we quote the nonresonant
cross sections with the respective $J/\psi$ and $\psi(2S)$ contributions 
excluded.
The quoted uncertainties are statistical and systematic, respectively.
The statistical uncertainty results from the statistical 
uncertainty in the number of selected $K^+K^-\gamma$ events. 
The systematic uncertainty includes the systematic uncertainty in the number
of events, the statistical uncertainty in the detection efficiency 
(1.5\%--4.0\%), and the uncertainties in the
efficiency correction (2.6\%--5.1\%),
the integrated luminosity (0.5\%)~\cite{lum}, and 
the ISR luminosity (0.5\%)~\cite{radf1,radf2}. 
For the mass interval 7.5--8.0 GeV/$c^2$, the 90\% confidence level (CL)
upper limit on the cross section is listed. The measured cross section is
shown in Fig.~\ref{figcs}.

It is more convenient to perform comparisons with previous measurements 
and theoretical predictions in terms of the form factor.
The values of the charged-kaon electromagnetic form factor 
obtained using Eq.~(\ref{eq4}) are listed in Table~\ref{sumtab}. 
The form factor is plotted in Fig.~\ref{fig19} as a function  of $M_{K^+K^-}$
over the range 2.6--5.0 GeV/$c^2$, together with all other
measurements~\cite{KKbabar,CLEO,NU}. The present measurement is consistent 
within the uncertainties with our previous, independent LA ISR 
result~\cite{KKbabar}, which used a data sample corresponding only to 
232 fb$^{-1}$, and provides a much better constraint on the mass dependence 
in this region.
\begin{figure}
\includegraphics[width=.4\textwidth]{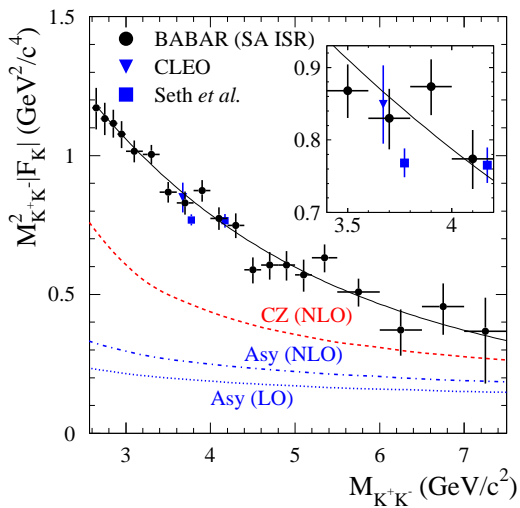}
\caption{(Color online) The scaled charged-kaon electromagnetic form factor 
measured
in this analysis [BABAR (SA ISR)] and in other experiments:
CLEO~\cite{CLEO} and Seth {\it et al}.~\cite{NU}. Our data are approximated by 
a smooth
curve. The dotted blue curve shows the LO pQCD prediction for
the form factor obtained with the asymptotic kaon distribution
amplitude. The dot-dashed blue and dashed red curves are the NLO pQCD predictions 
obtained with the asymptotic and Chernyak-Zhitnitsky distribution
amplitudes, respectively. The inset shows an enlarged version 
of the mass region 3.4--4.2 GeV/$c^2$.
\label{fig20}}
\end{figure}

To compare our results with the precise  
measurements of Refs.~\cite{CLEO,NU}, we plot
in Fig.~\ref{fig20} the scaled form factor $M_{K^+K^-}^2|F_K(M_{K^+K^-})|$ 
and fit our data with a smooth function $x^2|F_K(x)|=A/(x^\gamma+B)$,
where $A$, $B$, and $\gamma$ are fitted parameters. As seen in the inset, 
the CLEO and Seth {\it et al.} points at 3.67 and 4.17 GeV/$c^2$,
 respectively, are consistent with this function, whereas the Seth {\it et al.}
 point at 3.772 GeV/$c^2$ lies about three standard deviations below.
 Since this data point is obtained at
the maximum of the $\psi(3770)$ resonance, the deviation may be a
result of interference between the resonant and nonresonant amplitudes
of the $e^+e^-\to K^+K^-$ reaction, which we discuss in the next section.

The dotted curve in Fig.~\ref{fig20} represents the leading-order (LO)
asymptotic pQCD prediction of Eq.~(\ref{eqqcd}), calculated 
with $\alpha_s(M^2_{K^+K^-}/4)$~\cite{NLO}. It lies well below most 
of the data, which might be explained by higher-order pQCD, power 
corrections, and a deviation of the kaon distribution amplitude (DA),
which describes the quark relative momentum distribution
inside the meson, from its asymptotic shape.
The dot-dashed curve represents the leading-twist, 
next-to-leading-order (NLO) prediction using the asymptotic DA,
obtained by multiplying the pion form factor from Ref.~\cite{NLO} by a 
factor $f_K^2/f_\pi^2\approx 1.45$. The NLO correction leads
to an increase of about 20\%, nearly independent of mass. The dashed curve
represents the NLO prediction using the Chernyak-Zhitnitsky (CZ) DA,
obtained by multiplying the result in Ref.~\cite{NLO} by a factor of 
0.95~\cite{CZ}.

Our data lie well above all predictions.
However, they decrease faster than $\alpha_s/M_{K^+K^-}^2$ 
as $M_{K^+K^-}$ increases, and
are consistent with an approach to the pQCD prediction at higher mass.
 In particular, the ratio of the measured form factor to the
asymptotic pQCD prediction (curve Asy(LO) in Fig.~\ref{fig20}) changes from 
about 5.3 at 3 GeV$/c^2$ to about 2.6 at 7 GeV$/c^2$.

\section{ \boldmath $J/\psi$ and $\psi(2S)$ decays into $K^+K^-$}\label{psi}
\begin{figure*}
\includegraphics[width=.4\textwidth]{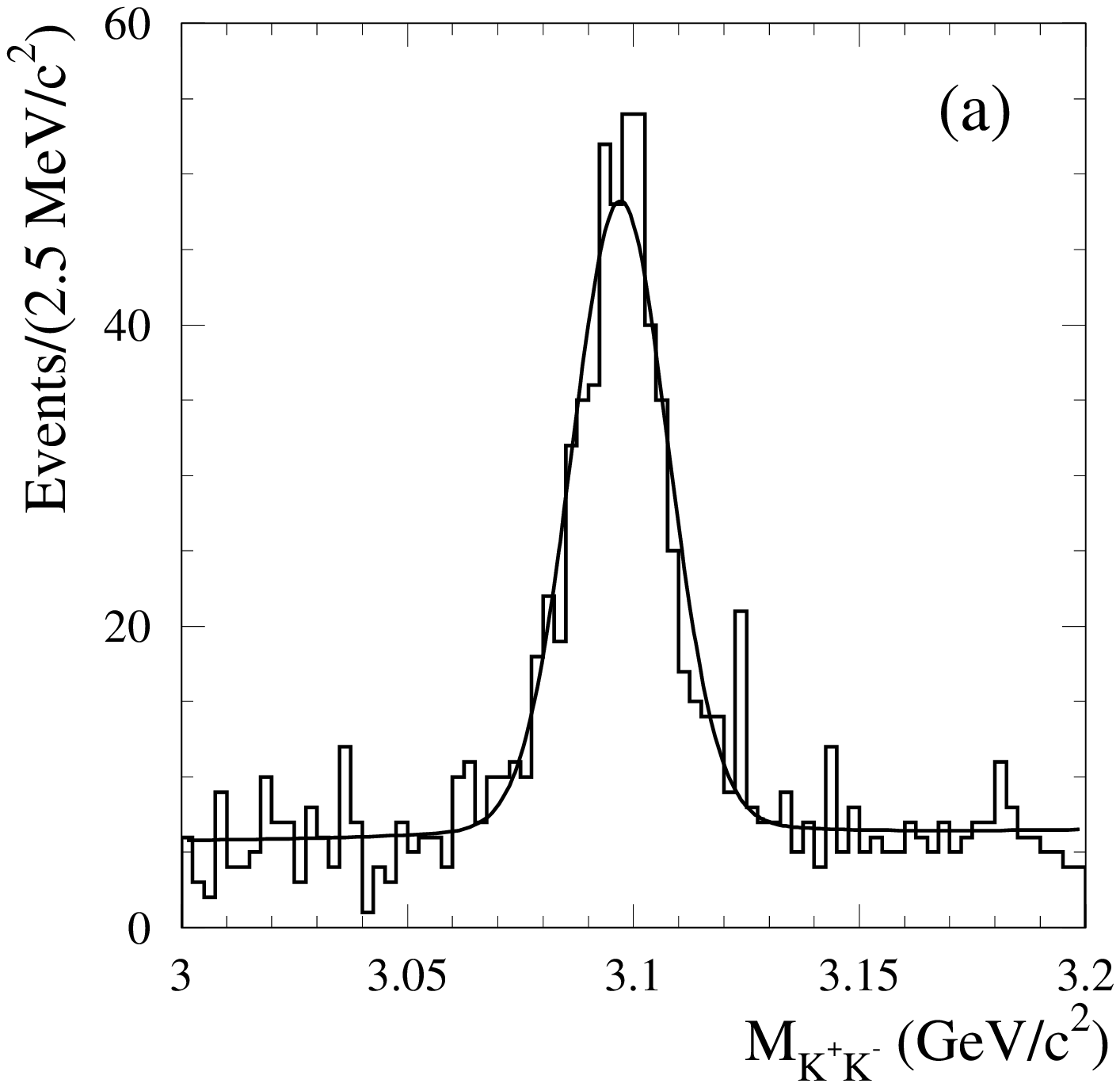}
\includegraphics[width=.4\textwidth]{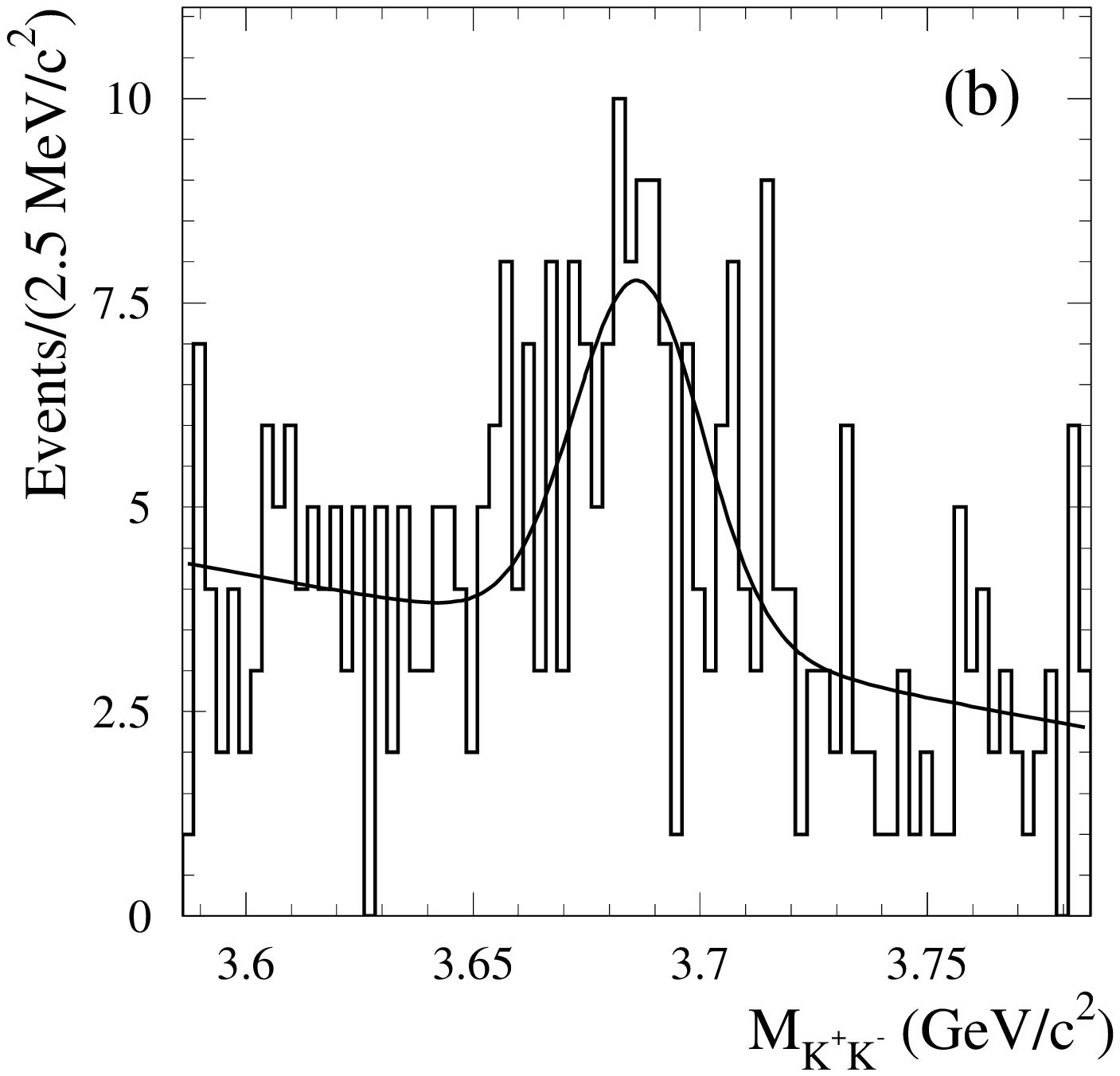}
\caption{The $K^+K^-$ mass spectra in the regions
near the $J/\psi$ (a) and $\psi(2S)$ (b) resonances. The curves 
exhibit the results of the fits described in the text.
\label{fig18}}
\end{figure*}

To study the production of $K^+K^-$ pairs through the $J/\psi$ and $\psi(2S)$
resonances, we increase the detection efficiency by
selecting events with the looser requirements
$p_{{\rm T},K^+K^-} < 1$ GeV/$c$ and $-2 < M_{\rm{miss}}^2 <3$ GeV$^2/c^4$.
The resulting $K^+K^-$ mass spectra in the $J/\psi$ and $\psi(2S)$
mass regions are shown in Fig.~\ref{fig18}.
Each of these spectra is fitted with the sum of a signal probability density 
function (PDF) and a linear background. The signal PDF is a 
Breit-Wigner (BW) function convolved with a double-Gaussian function
describing signal resolution. In each fit, the BW mass and width are fixed to
their known values~\cite{pdg} and the nominal resolution parameters are taken
from simulation. In order to account for deficiencies in the simulation, a
mass shift $\Delta M$ is allowed, and an increase in both Gaussian widths by
a term $\sigma_G$ added in quadrature is introduced. The free parameters in 
the
$J/\psi$ fit are the numbers of signal and background events, the slope
 of the background function, $\Delta M$, and $\sigma_G$. 
In the $\psi(2S)$ fit, $\sigma_G$ and $\Delta M$ are fixed to 
the values obtained for the $J/\psi$.

The fitted curves are shown in Fig.~\ref{fig18}. 
The numbers of $J/\psi$ and $\psi(2S)$ events are found to be
$462\pm28$ and $66\pm13$, respectively.
Results for other fitted parameters are $\sigma_G=3\pm3$ MeV/$c^2$ and
$\Delta M=M_{J/\psi}-M_{J/\psi}^{\rm MC}=(0.0\pm0.9)$ MeV/$c^2$.
The $\sigma_G$ value corresponds to a difference of about 4\% in mass 
resolution (11 MeV/$c^2$ at the $J/\psi$) between data and simulation.
The detection efficiencies, corrected for the data-MC simulation 
difference in detector response, 
are $(7.6\pm 0.2)\%$ for the $J/\psi$ and 
$(13.4\pm 0.3)\%$ for the $\psi(2S)$.

The total cross sections for the processes 
$e^+e^-\to \psi \gamma \to K^+K^-\gamma$, where 
$\psi$ is a narrow resonance like the $J/\psi$ or
$\psi(2S)$, is proportional to the electronic width of the resonance and 
its branching fraction into $K^+K^-$, i.e.,
$\sigma_{\psi\gamma}=a_\psi\Gamma(\psi\to e^+e^-)B(\psi\to K^+K^-)$. 
The coefficient
$a_\psi$ can be calculated by integrating Eq.~(\ref{eq1}) with $\sigma_{K^+K^-}$
set to the appropriate BW function. Using $W(s,x)$ from 
Refs.~\cite{radf1,radf2}, we obtain $a_{J/\psi}=6.91$ nb/keV 
and $a_{\psi(2S)}=6.07$ nb/keV. 
\begin{table*}
\caption{The products $\Gamma(\psi\to e^+e^-){\cal B}(\psi\to K^+K^-)$
and the branching fractions ${\cal B}(\psi\to K^+K^-)$ obtained in this work
for the $J/\psi$ and $\psi(2S)$ resonances. The directly measured values are
shown in the rows labeled ``Measured values''. In the rows marked 
as ``Corrected'',
the values of the products and branching fractions corrected for the shift
due to interference between resonant and nonresonant amplitudes are listed for
opposite signs of $\sin{\varphi}$. In the row marked ``$e^+e^-\to K^+K^-$ 
average'', we give the average of the values of the branching fractions 
measured in the reaction $e^+e^-\to K^+K^-$. 
In the row ``$\psi(2S)\to J/\psi \pi^+\pi^-$, $J/\psi \to K^+K^-$'',
the result for the $J/\psi \to K^+K^-$ branching fraction 
obtained Ref.~\cite{NUjpsi} is reported.
\label{tabpsi}}
\begin{ruledtabular}
\begin{tabular}{lcc}
$\Gamma(\psi\to e^+e^-){\cal B}(\psi\to K^+K^-)$ (eV) 
& $J/\psi$ & $\psi(2S)$ \\
\hline
\\[-2.1ex]
Measured value &
$1.86\pm 0.11\pm 0.05$ & $0.173\pm0.035\pm0.005$  \\
Corrected with $\sin{\varphi}>0$ &
$1.78\pm 0.11\pm 0.05$ & $0.147\pm0.035\pm0.005$\\
Corrected with $\sin{\varphi}<0$ &
$1.94\pm 0.11\pm 0.05$ & $0.197\pm0.035\pm0.005$\\
\hline
\\[-2.1ex]
${\cal B}(\psi\to K^+K^-)\times 10^4$ &
 $J/\psi$ & $\psi(2S)$ \\
\hline
\\[-2.1ex]
Measured value & $3.36\pm0.20\pm0.12$ & $0.73\pm0.15\pm0.02$\\
Corrected with $\sin{\varphi}>0$&$3.22\pm0.20\pm0.12$ & $0.62\pm0.15\pm0.02$\\
Corrected with $\sin{\varphi}<0$&$3.50\pm0.20\pm0.12$ & $0.83\pm0.15\pm0.02$\\
$e^+e^-\to K^+K^-$ average & $2.43\pm0.26$~\cite{jpsikk1,jpsikk2,KKbabar} & $0.71\pm0.05$~\cite{pdg}\\
$\psi(2S)\to J/\psi \pi^+\pi^-$, $J/\psi \to K^+K^-$~\cite{NUjpsi}& $2.86\pm0.21$ & \\
\end{tabular}
\end{ruledtabular}
\end{table*}

From the measured values of the cross sections, 
$\sigma_{\psi\gamma}^{\rm exp}=N_{\psi}/(\varepsilon L)$, we obtain the 
measured values 
of the products $\Gamma(\psi\to e^+e^-){\cal B}(\psi\to K^+K^-)$ listed
in Table~\ref{tabpsi}. The term ``Measured value'' is used because
the value of the product obtained in this way may differ from the true 
value due to interference with the nonresonant process $e^+e^- \to K^+K^-$,
as discussed below. 
The quoted systematic uncertainty includes the uncertainties
in the detection efficiency, the integrated luminosity (0.5\%), and the 
theoretical uncertainty in the ISR luminosity (0.5\%).
 Using the nominal values of the electronic widths~\cite{pdg},
$\Gamma(J/\psi\to e^+e^-)=5.55\pm0.14$ keV
and $\Gamma(\psi(2S)\to e^+e^-)=2.37\pm0.04$ keV, we calculate 
the measured values of the $\psi\to K^+K^-$ branching fractions 
listed in Table~\ref{tabpsi}.
Since the decay $\psi(2S)\to K^+K^-$ was studied previously only 
in the reaction $e^+e^-\to K^+K^-$, our measurement of 
${\cal B}(\psi(2S)\to K^+K^-)$ can be directly  compared with the PDG 
value~\cite{pdg}, $(0.71\pm 0.05)\times 10^{-4}$. Although it is less precise,
our measured value agrees well with that from Ref.~\cite{pdg}.
For $J/\psi\to K^+K^-$ there are several 
measurements~\cite{jpsikk1,jpsikk2,KKbabar} in the $e^+e^-\to K^+K^-$ reaction,
and one measurement~\cite{NUjpsi} in which $J/\psi$'s were produced in the 
$\psi(2S)\to J/\psi\pi^+\pi^-$ decay. To compare with the $e^+e^-$ 
measurements, we calculate the average of the 
results~\cite{jpsikk1,jpsikk2,KKbabar}, and obtain the value
$(2.43\pm0.26)\times 10^{-4}$. 
Our result is larger than this average by 2.7
standard deviations. A comparison with the measurement of Ref.\cite{NUjpsi} 
is presented below after applying a correction for interference.

To estimate the effect of interference, we represent 
the c.m. energy ($E$) dependence of the $e^+e^- \to K^+K^-$ cross section  
near the $\psi$ resonance as~\cite{gatto} 
\begin{widetext}
\begin{multline}
\sigma_{K^+K^-}(E)=\sigma_0
\left|1-\sqrt{\frac{\sigma_\psi}{\sigma_0}}(A_\gamma+A_se^{i\varphi})
\frac{m\Gamma}{D}\right|^2=\\
\sigma_0+\sigma_\psi \left [{\cal B}(\psi\to K^+K^-)+
2\sqrt{\frac{\sigma_0}{\sigma_\psi}}A_s\sin{\varphi}\right]
\frac{m^2\Gamma^2}{|D|^2}-
2\sqrt{\sigma_0\sigma_\psi}(A_\gamma+A_s\cos{\varphi})
\frac{m\Gamma(m^2-E^2)}{|D|^2},
\label{inter1}
\end{multline}
\end{widetext}
where $\sigma_0$ is the nonresonant cross section [Eq.~(\ref{eq4})],
$\sigma_\psi=(12\pi/m^2){\cal B}(\psi\to e^+e^-)$,
$A_\gamma$ and $A_s$ are the moduli of the single-photon and strong
$\psi$ decay amplitudes, respectively, $\varphi$ is their relative phase,
$D=m^2-E^2-im\Gamma$, and $m$ and $\Gamma$ 
are the resonance mass and width, respectively.
The decay amplitudes are defined such that
${\cal B}(\psi\to K^+K^-)=|A_\gamma+A_se^{i\varphi}|^2$.
The value of the single-photon contribution is related to the
kaon form factor through
\begin{equation}
A_\gamma^2={\cal B}(\psi\to e^+e^-)\frac{|F_K(m)|^2}{4}\beta^3(m),
\end{equation}
where $\beta$ is the phase-space factor from Eq.~(\ref{eq4}).

For narrow resonances, the interference term proportional to 
$m^2-E^2$ integrates to zero due to the beam energy spread in direct
$e^+e^-$ experiments and detector resolution in ISR measurements.
The remaining interference term has a BW shape and  
causes a shift of the measured  
${\cal B}(\psi\to K^+K^-)$ relative to its true value by
\begin{equation}
\delta{\cal B}=2\sqrt{\frac{\sigma_0}{\sigma_\psi}}A_s\sin{\varphi}.
\label{dB}
\end{equation}

The values of $\cos{\varphi}$ and $A_s$ can be obtained from
a combined analysis of the $\psi\to K^+K^-$ and $\psi\to K_SK_L$ decays,
whose branching fractions depend on the same strong amplitude~\cite{ph1,ph2}
\begin{eqnarray}
{\cal B}(\psi\to K^+K^-)&=&\left |A_\gamma^{K^+K^-}+A_se^{i\varphi} \right |^2,
\nonumber\\
{\cal B}(\psi\to K_SK_L)&=&\left |\kappa A_\gamma^{K^+K^-}+A_se^{i\varphi} \right |^2,
\label{eqbr}
\end{eqnarray}
where $\kappa$ is the ratio of the single-photon amplitudes for
the $\psi\to K_SK_L$ and $\psi\to K^+K^-$ decays, and
$|\kappa|=A_\gamma^{K_SK_L}/A_\gamma^{K^+K^-}$.
It is expected that in the energy region under study, the single-photon 
amplitudes have the same sign of the real parts, and similar ratios of 
the imaginary-to-real parts~\cite{chernyak_sign},
i.e., $\kappa$ is a positive real number to a good approximation.

For the branching fraction ${\cal B}(J/\psi\to K^+K^-)$ in Eqs.~(\ref{eqbr}),
we use the average of the existing 
measurements~\cite{jpsikk1,jpsikk2,KKbabar,NUjpsi} and our result.
For ${\cal B}(J/\psi\to K_SK_L)$ there are two relatively precise measurements,
$(1.82\pm0.14)\times10^{-4}$~\cite{BESKSKL}
and $(2.62\pm0.21)\times10^{-4}$~\cite{NUjpsi}, 
which are not consistent with each other. We solve 
Eqs.~(\ref{eqbr}) separately for these two values of 
${\cal B}(J/\psi\to K_SK_L)$.   
The branching fractions measured in the reactions
$e^+e^-\to K^+K^-$ and $e^+e^-\to K_SK_L$ are corrected as
${\cal B}\to{\cal B}-\delta {\cal B}$ before averaging, 
where $\delta {\cal B}$ is given by Eq.~(\ref{dB}).
This correction is not needed for the measurements of Ref.~\cite{NUjpsi}, in 
which the $J/\psi$ mesons are produced in $\psi(2S)\to J/\psi \pi^+\pi^-$ decays.
For $\psi(2S)$ decays we use the branching fraction values from 
Ref.~\cite{pdg} corrected using Eq.~(\ref{dB}).

The coefficient $|\kappa|$ in Eqs.~(\ref{eqbr}) is equal to the ratio
of the neutral- and charged-kaon form factors $|F_{K^0}/F_K|$.
Data on $F_{K^0}$ above 2 GeV are scarce. There are 
two measurements~\cite{dm1,KSKLbabar} near 2 GeV, from which we
estimate $|\kappa|=0.28\pm0.08$, and there is only one measurement at higher
energy, namely $|\kappa|=0.12\pm 0.04$ at 4.17 GeV~\cite{NUkskl}.  
Using linear interpolation we estimate $|\kappa|=0.2\pm 0.1$ at 
the mass of the $J/\psi$
and $0.15\pm 0.07$ at the mass of the $\psi(2S)$.
The values of the charged-kaon form factor,
$F_K(M_{J/\psi})=0.107\pm0.002$ and $F_K(M_{\psi(2S)})=0.0634\pm0.0014$,
needed to calculate $A_\gamma^{K^+K^-}$, are taken from 
the fit to our form factor data shown in Fig.~\ref{fig20}.

The values of $\varphi$ and $\delta{\cal B}(\psi\to K^+K^-)$
obtained using Eqs.~(\ref{dB}) and (\ref{eqbr}) are 
listed in Tables~\ref{psi1phase} and \ref{psi2phase}. 
Since Eqs.~(\ref{eqbr})
do not allow us to determine the sign of $\sin{\varphi}$, 
the calculations are 
performed twice, once assuming $\sin{\varphi}>0$ and 
once assuming $\sin{\varphi}<0$.
For the results in Table~\ref{psi1phase} the two upper (bottom) rows marked 
``BES'' (``Seth {\it et al.}'') present results obtained using 
${\cal B}(J/\psi\to K_SK_L)$
from Ref.~\cite{BESKSKL} (Ref.~\cite{NUjpsi}).
\begin{table*}
\caption{The relative phase ($\varphi$) between the single-photon and 
strong amplitudes for $J/\psi\to K \bar{K}$ decays calculated with 
$\kappa=A_\gamma^{K_SK_L}/A_\gamma^{K^+K^-}=0.2\pm0.1$ and $\kappa=0$,
and the correction to the value of ${\cal B}(J/\psi\to K^+K^-)$ measured 
in the reaction $e^+e^-\to K^+K^-$. The calculation is performed for 
the value of ${\cal B}(J/\psi\to K_SK_L)$ obtained in Ref.~\cite{BESKSKL},
the value obtained in Ref.~\cite{NUjpsi}, and assuming either a positive or 
negative value for $\sin{\varphi}$.
\label{psi1phase}}
\begin{ruledtabular}
\begin{tabular}{cccc}
$J/\psi\to K_SK_L$ & $\varphi$ & $\varphi(\kappa=0)$ &
 $\delta{\cal B}(J/\psi\to K^+K^-)\times 10^4$ \\
\hline
\\[-2.1ex]
BES~\cite{BESKSKL} & $(97\pm 5)^\circ$ & $(98\pm4)^\circ$ & $0.13\pm0.01$ \\
                   & $-(97\pm 5)^\circ$ & $-(96\pm4)^\circ$ & $-0.13\pm0.01$ \\
Seth {\it et al.}~\cite{NUjpsi}   & $(111\pm5)^\circ$& $(108\pm4)^\circ$ & $0.15\pm0.01$ \\
                   & $-(109\pm 5)^\circ$ & $-(107\pm4)^\circ$ & $-0.15\pm0.01$ \\
\end{tabular}
\end{ruledtabular}
\end{table*}
\begin{table}
\caption{The relative phase ($\varphi$) between the single-photon and 
strong amplitudes for $\psi(2S)\to K \bar {K}$ decays calculated with 
$\kappa=A_\gamma^{K_SK_L}/A_\gamma^{K^+K^-}=0.15\pm0.07$ and $\kappa=0$,
and the correction to the value of ${\cal B}(\psi(2S)\to K^+K^-)$ 
measured in the reaction $e^+e^-\to K^+K^-$. The calculation was performed 
for each sign of $\sin{\varphi}$.
\label{psi2phase}}
\begin{ruledtabular}
\begin{tabular}{ccc}
$\varphi$ & $\varphi (\kappa=0)$ &
 $\delta{\cal B}(\psi(2S)\to K^+K^-)\times 10^4$ \\
\hline
\\[-2.1ex]
$(82\pm 12)^\circ$ & $(92\pm9)^\circ$ & $0.11\pm0.01$ \\
$-(58\pm 14)^\circ$ & $-(57\pm12)^\circ$ & $-0.10\pm0.02$ \\
\end{tabular}
\end{ruledtabular}
\end{table}
We also list the results obtained for $\kappa=0$,
corresponding to the  assumption $A_\gamma^{K_SK_L}\ll A_\gamma^{K^+K^-}$ 
used for most previous determinations of $\varphi$. It is seen that 
allowing $A_\gamma^{K_SK_L}$ to be non-zero does not lead to a significant
change in the results. 

For the $J/\psi$, for which the most precise measurement of 
${\cal B}(J/\psi\to K^+K^-)$ 
was performed in $\psi(2S)$ decay, the result for $\cos{\varphi}$ is weakly
dependent on the sign of $\sin{\varphi}$. We confirm the conclusion of
Refs.~\cite{ph1,ph2} to the effect that the strong amplitude describing
$J/\psi\to K^+K^-$ decay
has a large imaginary part. Using ${\cal B}(J/\psi\to K_SK_L)$ from 
Ref.~\cite{NUjpsi}, a non-negligible real part of the strong amplitude is
obtained. 

For the $\psi(2S)$, the result on $\cos{\varphi}$ is strongly dependent
on the sign of $\sin{\varphi}$. Here theoretical arguments may help
to choose the sign. The ratio of the strong amplitudes 
for $\psi(2S)\to K\bar{K}$ and $J/\psi\to K\bar{K}$ decays 
is expected~\cite{12rule} to be
\begin{multline}
\frac{A_s^2(\psi(2S)\to K\bar{K})}{A_s^2(J/\psi\to K\bar{K})}\approx\\
\frac{{\cal B}(\psi(2S)\to e^+e^-)}{{\cal B}(J/\psi\to e^+e^-)}
\frac{\beta^3(M_{\psi(2S)})}{\beta^3(M_{J/\psi})}=0.138\pm0.003.
\end{multline}
The experimental value of this ratio (for $J/\psi$ we used $A_s$ obtained 
with ${\cal B}(J/\psi\to K_S K_L)$ from Ref.~\cite{NUjpsi}) is 
$0.192\pm0.026$ for $\sin{\varphi}<0$ and $0.170\pm0.023$ for 
$\sin{\varphi}>0$. The result for the positive sign is in slightly better
agreement with the prediction.

Using the values of $\delta{\cal B}$ given in Tables ~\ref{psi1phase} and 
\ref{psi2phase}, we correct the measured values of the products
$\Gamma(\psi\to e^+e^-){\cal B}(\psi\to K^+K^-)$ and the branching fractions
and list the corrected values in Table~\ref{tabpsi}. 

The corrected values of ${\cal B}(J/\psi\to K^+K^-)$ can be compared with the 
measurement of Ref.~\cite{NUjpsi} $(2.86\pm 0.21)\times 10^{-4}$.
The difference between the two measurements is $2\sigma$ for $\sin{\varphi}<0$,
and $1\sigma$ for $\sin{\varphi}>0$. Our result for $J/\psi\to K^+K^-$ 
thus provides an indication that $\sin{\varphi}$ is positive. 
It should be stressed that the shifts we find between the measured and true 
values of ${\cal B}(\psi\to K^+K^-)$ are significant: about 5\% for the 
$J/\psi$ and about 15\% for the $\psi(2S)$.
Thus the interference effect should be taken into account in any future 
precise measurements
of the branching fraction for $J/\psi\to K^+K^-$ or $\psi(2S)\to K^+K^-$.

\section{Summary} \label{summary}
The process $e^+e^-\to K^+K^-\gamma$ has been studied in
the $K^+K^-$ invariant mass range from 2.6 to 8 GeV/$c^2$ 
using events in which the photon is emitted close to the collision axis.
From the measured $K^+K^-$ mass spectrum we obtain the
$e^+e^-\to K^+K^-$ cross section and determine the charged-kaon
electromagnetic form factor (Table~\ref{sumtab}). 
This is the first measurement of
the kaon form factor for $K^+K^-$ invariant masses higher
than 5 GeV/$c^2$ and the most precise measurement in the range 
2.6--5.0 GeV/$c^2$. Our data  indicate clearly that the difference between 
the measured form factor and the leading twist pQCD prediction decreases 
with increasing $K^+K^-$ invariant mass.

We present measurements of the $J/\psi\to K^+K^-$ and $\psi(2S)\to K^+K^-$
branching fractions (Table~\ref{tabpsi}).
Using the measured values
of the branching fractions and charged-kaon form factors, and data
from other experiments on $e^+e^-\to K_S K_L$ and $\psi\to K\bar{K}$ decays,
we have determined the phase difference $\varphi$ between the strong and 
single-photon amplitudes for $J/\psi\to K\bar{K}$ and 
$\psi(2S)\to K\bar{K}$ decays. We have calculated the shifts in the measured
values of the branching fractions due to interference between resonant and
nonresonant amplitudes in the $e^+e^-\to K^+K^-$ reaction. The shift has been 
found to be relatively large, about $\pm 5\%$ for the $J/\psi$ and about 
$\pm 15\%$ for the $\psi(2S)$, where the sign is determined by the sign
of $\sin{\varphi}$. 

It should be noted that the sign of $\sin{\varphi}$ for $J/\psi$ decays can be 
determined experimentally from the difference, $\delta{\cal B}/{\cal B}$, 
between the $J/\psi\to K^+K^-$ branching fractions measured in the reaction 
$e^+e^-\to K^+K^-$ and in the decay $\psi(2S)\to J/\psi \pi^+\pi^-$.
We hope that this measurement will be performed in future experiments.

\section{ \boldmath Acknowledgments}
We thank V.~L.~Chernyak for useful discussions.
We are grateful for the
extraordinary contributions of our \pep2\ colleagues in
achieving the excellent luminosity and machine conditions
that have made this work possible.
The success of this project also relies critically on the
expertise and dedication of the computing organizations that
support \babar.
The collaborating institutions wish to thank
SLAC for its support and the kind hospitality extended to them.
This work is supported by the
US Department of Energy
and National Science Foundation, the
Natural Sciences and Engineering Research Council (Canada),
the Commissariat \`a l'Energie Atomique and
Institut National de Physique Nucl\'eaire et de Physique des Particules
(France), the
Bundesministerium f\"ur Bildung und Forschung and
Deutsche Forschungsgemeinschaft
(Germany), the
Istituto Nazionale di Fisica Nucleare (Italy),
the Foundation for Fundamental Research on Matter (The Netherlands),
the Research Council of Norway, the
Ministry of Education and Science of the Russian Federation,
Ministerio de Econom\'{\i}a y Competitividad (Spain), the
Science and Technology Facilities Council (United Kingdom),
and the Binational Science Foundation (U.S.-Israel).
Individuals have received support from
the Marie-Curie IEF program (European Union) and the A. P. Sloan 
Foundation (USA).

\end{document}